\documentclass[aps,pre,twocolumn,showpacs,preprintnumbers,a4paper]{revtex4} 

\usepackage{graphics}
\usepackage{epsf}
\usepackage{epsfig}

\begin{document} 


\title{Universal finite-size scaling behavior 
and universal dynamical scaling behavior of
absorbing phase transitions with a conserved field}

\author{S. L\"ubeck}
\affiliation{Weizmann Institute of Science, 
Department of Physics of Complex Systems, 
76100 Rehovot, Israel,
}

\author{P.\,C.~Heger}
\affiliation{
Institut f\"ur Theoretische Physik,
Universit\"at Duisburg-Essen, 
47048 Duisburg, Germany}

\date\today
\date{July 13, 2003}

\begin{abstract}
We analyze numerically three different models
exhibiting an absorbing phase transition.
We focus on the finite-size scaling as well as the
dynamical scaling behavior.
An accurate determination of several critical 
exponents allows to validate certain
hyperscaling relations.
Using this hyperscaling relations it is 
possible to express the 
avalanche exponents of a self-organized 
critical system in terms of the ordinary exponents
of a continuous absorbing phase transition.
\end{abstract}



\preprint{accepted for publication in {\it Physical Review E} 2003}

\maketitle

\section{Introduction}

Absorbing phase transition (APT) are a particular class
of non-equilibrium phase transitions occurring
in physical, biological, as well as chemical 
systems (see for instance~\cite{HINRICHSEN_1}).
Transitions to absorbing states are of particular
interest since they have no equilibrium counterparts
and may occur even in one-dimensional systems.
A characteristic feature of absorbing phase transitions
is the competition between the proliferation and 
annihilation of a certain entity~$A$, e.g., particles,
energy units, viruses, molecules in catalytic reactions, etc.
It is essential that no spontaneous creation of such 
quantities takes place.
At a critical value of the proliferation-annihilation rate
the density $\rho(A)$ vanishes and the system
is trapped forever in the absorbing state $\rho(A)=0$.

Directed percolation is recognized as a paradigmatic
example of absorbing phase transitions.
This is reflected by the universality hypothesis of Janssen
and Grassberger that models which exhibit a continuous
phase transition to a single absorbing state generally
belong to the universality class of directed 
percolation~\cite{JANSSEN_1,GRASSBERGER_2}.
Different universality classes occur for instance 
in the presence of additional symmetries.
In particular, particle conservation may lead to
the different universality class of absorbing phase 
transitions with a conserved field as pointed 
out in~\cite{ROSSI_1}.
For instance the conserved lattice gas (CLG)~\cite{ROSSI_1},
the conserved threshold transfer process (CTTP)~\cite{ROSSI_1},
the well known Manna sandpile model~\cite{MANNA_2},
as well as a reaction-diffusion model~\cite{PASTOR_2}
belong to this universality class~\cite{LUEB_26}.
Note that this universality class is of particular interest
since the corresponding systems connect the critical
behavior of the absorbing phase transition with the 
critical steady state of self-organized 
critical (SOC) systems~\cite{BAK_1}.
Actually, SOC sandpile models can be
considered as driven-dissipative versions of 
(closed) systems exhibiting absorbing phase 
transitions~\cite{VESPIGNANI_4}.

In this paper we consider the universal finite-size scaling 
as well as the 
universal dynamical scaling behavior of the CLG model, 
the CTTP, and the Manna model.
First we introduce a method that allows to
study finite-size effects in the steady state.
In contrast to previous attempts to measure finite-size
effects our method is well defined.
Furthermore it can be applied immediately to 
other classes of absorbing phase transitions.
Second we consider the activity spreading 
of a single active seed.
The corresponding spreading exponents are naturally 
connected to the avalanche exponents of SOC 
systems~\cite{MUNOZ_3}.
In particular we discuss certain hyperscaling laws
relating the spreading exponents to the steady state 
exponents of absorbing phase transitions.
This allows us to express the
SOC avalanche exponents
in terms of the exponents of the corresponding 
absorbing phase transition
(e.g. the exponents of the order parameter, the correlation length
exponent, etc.).
Thus the critical state of SOC systems is closely related
to the critical properties of an ordinary second order phase 
transition.

\section{Models}

The first considered model is the conserved lattice 
gas (CLG)~\cite{ROSSI_1} which is a stochastic variant
of a model introduced by Jensen~\cite{JENSEN_7}.
In the CLG model lattice sites may be empty or occupied
by one particle.
In order to mimic a repulsive interaction a given particle
is considered as active if at least one of its
neighboring sites on the lattice is occupied by another
particle.
If all neighboring sites are empty the particle remains
inactive.
Active particles are moved in the next update
step to one of their empty nearest neighbor sites,
selected at random.

The second model is the so-called conserved threshold transfer
process (CTTP)~\cite{ROSSI_1},
a modification of the threshold transfer process
introduced in~\cite{MENDES_1}.
Here, lattice sites may be empty, occupied by one particle,
or occupied by two particles.
Empty and singly occupied sites are considered as
inactive whereas double occupied lattice sites are 
considered as active.
In the latter case one tries to transfer both particles
of a given active site to randomly chosen empty or singly
occupied nearest neighbor sites.

The third model is a modified version of the
Manna sandpile model~\cite{MANNA_2} the so-called fixed-energy
Manna model~\cite{VESPIGNANI_4}.
In contrast to the CTTP the Manna model allows unlimited
particle occupation of lattice sites.
We use in our investigations the original Manna relaxation
rules, i.e., 
lattice sites which are occupied by at least two particles
are considered as active and all 
particles are moved to the neighboring
sites selected at random.

\begin{figure}[t]
  \includegraphics[width=7.0cm,angle=0]{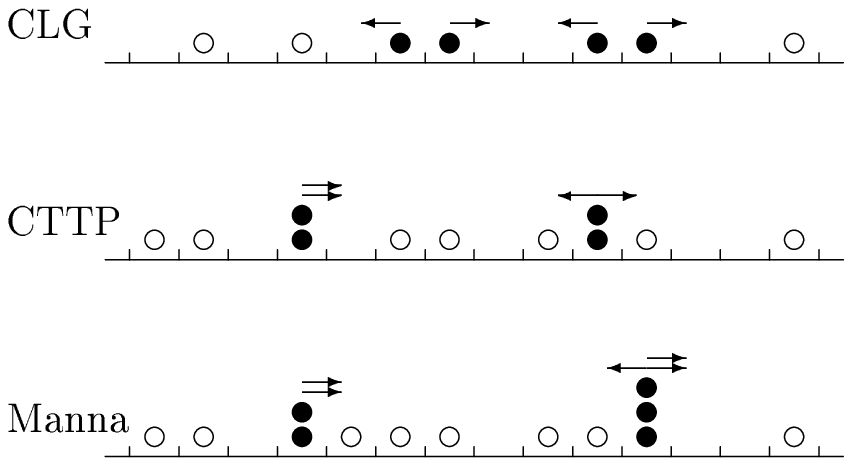}
  \caption{
  Sketch of the dynamics of the three considered models.
  Filled circles mark active particles 
  whereas non-active particles are marked by open circles.  
  The arrows denote how the active particles are (probably) 
  moved in the next update step. 
  In the case of the one-dimensional CLG model
  the particle transfer is deterministic.
  For the one-dimensional CTTP stochastic (left) 
  as well as deterministic (right) particle movements
  may occur.
  Only the one-dimensional Manna model is characterized by 
  a full stochastic dynamics in $D=1$.}
  \label{fig:sketch_models_01}
\end{figure}

The three models are sketched 
in Fig.\,\ref{fig:sketch_models_01}.
We use in all cases periodic boundary conditions,
i.e., closed systems are considered and the 
number of particles is conserved.
In our simulations (see~\cite{LUEB_22,LUEB_24} for 
details) we start from a random distribution 
of particles.
All models reach after a transient 
regime a steady state which is
characterized by the average density of active 
sites~$\rho_{\scriptscriptstyle \text a}$.
The density~$\rho_{\scriptscriptstyle \text a}$ is the order parameter
and the particle density~$\rho$ is the control parameter
of the absorbing phase transition, i.e., the order parameter
vanishes at the critical density~$\rho_{\text c}$ according to
$\rho_{\scriptscriptstyle \text a} \propto \delta\rho^{\beta}$,
with the reduced control parameter 
$\delta\rho=\rho/\rho_{\text c}-1$. 
Additionally to the order parameter we consider its 
fluctuations $\Delta \rho_{\scriptscriptstyle \text a}$.
Approaching the transition point from above ($\delta\rho>0$) 
the fluctuations diverge according to 
$\Delta \rho_{\scriptscriptstyle \text a}\propto  \delta\rho^{-\gamma^{\prime}}$
(see~\cite{LUEB_22,LUEB_24}).
Below the critical density (in the absorbing state)
the order parameter as well as its fluctuations
are zero in the steady state.

Similar to equilibrium phase transitions it is 
possible in the case of absorbing phase transitions
to apply an external field~$h$ which is
conjugated to the order parameter, i.e., the field  
causes a spontaneous creation of active particles
(see for instance~\cite{HINRICHSEN_1}).
A realization of the external field 
for absorbing phase transitions with a conserved field
was developed in~\cite{LUEB_22} where the
external field triggers movements of inactive
particles which may be activated in this way.
At the critical density~$\rho_{\scriptscriptstyle \text c}$
the order parameter and its fluctuations scale as
$\rho_{\scriptscriptstyle \text a} \propto h^{\beta/\sigma}$ 
and 
$\Delta\rho_{\scriptscriptstyle \text a} \propto h^{-\gamma^{\prime}/\sigma}$, 
respectively.

It was shown recently that the order parameter and its 
fluctuations obey in the steady state 
the scaling forms~\cite{LUEB_26}
\begin{eqnarray}
\label{eq:scal_ansatz_EqoS}
\rho_{\scriptscriptstyle \text a}(\delta\rho, h) 
\; & \sim & \; 
\lambda^{-\beta}\, \, {\tilde R}
(a_{\scriptscriptstyle \rho}  
\delta \rho \; \lambda, a_{\scriptscriptstyle h} h \;
\lambda^{\sigma}) \, ,\\
\label{eq:scal_ansatz_Fluc}
a_{\scriptscriptstyle \Delta} \,
\Delta \rho_{\scriptscriptstyle \text a}(\delta\rho, h) 
\; & \sim & \; 
\lambda^{\gamma^{\prime}}\, \, {\tilde D}
(a_{\scriptscriptstyle \rho} \delta \rho \; \lambda, 
a_{\scriptscriptstyle h} h \, \lambda^{\sigma})  \, .
\end{eqnarray}
The universal scaling functions ${\tilde R}(x,y)$ and 
${\tilde D}(x,y)$ are the same for all systems belonging 
to a given universality class whereas all non-universal 
system-dependent features (e.g.~the lattice structure, 
the update scheme, etc.)
are contained in the so-called non-universal metric factors 
$a_{\scriptscriptstyle \rho}$, $a_{\scriptscriptstyle h}$,
and $a_{\scriptscriptstyle \Delta}$~\cite{PRIVMAN_3}.
The universal scaling functions are normed
by the conditions ${\tilde R}(1,0)={\tilde R}(0,1)={\tilde D}(0,1)=1$
and the non-universal metric factors can be determined
from the amplitudes of 
\begin{eqnarray}
\label{eq:metric_factors_a_rho}
\rho_{\scriptscriptstyle \text a}(\delta \rho, h=0) \; & \sim & \; 
(a_{\scriptscriptstyle \rho} \, \delta \rho)^{\beta} \, ,\\
\label{eq:metric_factors_a_h}   
\rho_{\scriptscriptstyle \text a}(\delta \rho =0, h) \; & \sim & \; 
(a_{\scriptscriptstyle h} \, h)^{\beta / \sigma} \, , \\
\label{eq:metric_factors_a_Delta}   
a_{\scriptscriptstyle \Delta} \,
\Delta\rho_{\scriptscriptstyle \text a}(\delta \rho=0, h) \; & \sim &\; 
(a_{\scriptscriptstyle h} \, h)^{-\gamma^{\prime}/\sigma} \, .
\end{eqnarray}
These equations are obtained by choosing
in the scaling forms [Eqs.\,(\ref{eq:scal_ansatz_EqoS},
\ref{eq:scal_ansatz_Fluc})] 
$a_{\scriptscriptstyle \rho} \delta\rho \, \lambda=1$ 
and $a_{\scriptscriptstyle h} h \, \lambda^{\sigma}=1$, respectively.

\begin{figure}[b]
  \includegraphics[width=7.0cm,angle=0]{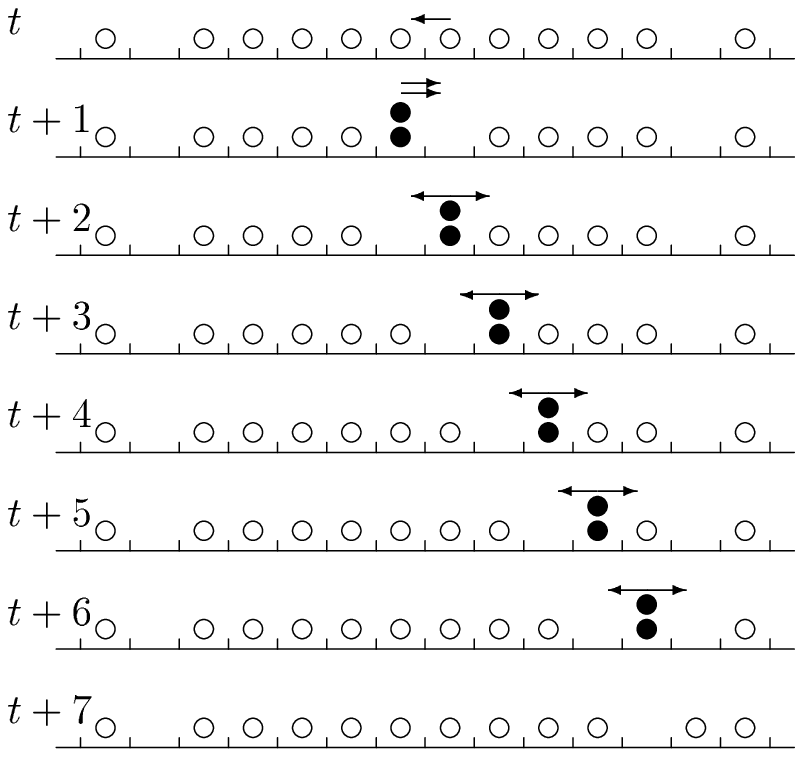}
  \caption{
  Sketch of the dynamics of the one-dimensional CTTP.
  At update step $t$ an active site is triggered by the
  external field in a cluster of inactive sites.
  The arrows denote how the active particles (full circles) are 
  moved in the next update step. 
  Due to the dynamic rules of the CTTP only two
  different relaxation processes (selected randomly) 
  occur: both particles
  are moved to the same empty site or both particles
  are moved to two adjacent sites.
  Thus the dynamics of the one-dimensional CTTP is 
  characterized by a trivial random walk of the perturbation.
  This random walk proceeds until it reaches the boundary
  of a cluster. }
  \label{fig:sketch_cttp_01}
\end{figure}

A recently performed analysis of the universal scaling functions
as well as of the critical exponents for $D\ge 2$~\cite{LUEB_26}
confirms the conjecture of~\cite{ROSSI_1} that the CLG model,
the CTTP, and the Manna model belong to the same universality
class. 
The situation is more complicated in one-dimensional
systems where a splitting of the universality class occur.
The reason for the non-universal behavior is that the 
dimensional reduction changes the stochastic character of the 
dynamics (see Fig.\,\ref{fig:sketch_models_01}.).
For instance the CLG model is characterized by deterministic
toppling rules in $D=1$ and exhibits a trivial
phase transition with $\beta=1$ 
and $\rho_{\scriptscriptstyle \text c}=1/2$ (see also
\mbox{ref.\,22} in~\cite{ROSSI_1}).
Due to the trivial behavior of the one-dimensional
CLG model we consider in this work the two- and 
three-dimensional CLG model only.

In the case of the CTTP we observe that roughly 4$0\%$ of the 
relaxation events are deterministic.
Furthermore a perturbation that is triggered by the
external field performs a simple random walk 
(see Fig.\,\ref{fig:sketch_cttp_01}).
This pathologic behavior is completely different
from the behavior of the one-dimensional Manna 
model that is characterized by a pure stochastic
relaxation of active particles to the next neighbors.
More than the other models the Manna model is therefore 
the paradigm of the universality class of absorbing phase 
transitions with a conserved field.
In the following we will call that class the Manna 
universality class 
since universality classes are often labeled by the 
simplest model belonging to them.

The universality splitting of the one-dimensional systems 
is in full agreement with the 
universality hypothesis of sandpile models~\cite{BENHUR_1}.
According to this conjecture the universality classes 
of sandpile models are determined by the way the
particles are distributed to the next neighbors
(deterministic, stochastic, directed, undirected, etc.).
Obviously the Manna universality class is characterized
by a stochastic and undirected distribution of 
particles.

\section{Steady-state finite-size scaling}

Similar to equilibrium critical phenomena we assume that the
system size~$L$ enters the scaling forms 
[Eqs.\,(\ref{eq:scal_ansatz_EqoS},\ref{eq:scal_ansatz_Fluc})] 
as an additional scaling field, i.e. 
\begin{eqnarray}
\label{eq:scal_ansatz_EqoS_FSS}
\rho_{\scriptscriptstyle \text a}(\delta\rho, h, L) 
& \sim & 
\lambda^{-\beta} {\tilde R}_{\scriptscriptstyle \text {pbc}}
(a_{\scriptscriptstyle \rho}  
\delta \rho  \lambda, a_{\scriptscriptstyle h} h 
\lambda^{\sigma}, 
a_{\scriptscriptstyle L} L \lambda^{-\nu_{\scriptscriptstyle \perp}}) \, ,\nonumber\\
\label{eq:scal_ansatz_Fluc_FSS}
a_{\scriptscriptstyle \Delta} 
\Delta \rho_{\scriptscriptstyle \text a}(\delta\rho, h, L) 
& \sim & 
\lambda^{\gamma^{\prime}} {\tilde D}_{\scriptscriptstyle \text {pbc}}
(a_{\scriptscriptstyle \rho} \delta \rho  \lambda, 
a_{\scriptscriptstyle h} h  \lambda^{\sigma} ,
a_{\scriptscriptstyle L} L \lambda^{-\nu_{\scriptscriptstyle \perp}}) ,
\end{eqnarray}
where the exponent $\nu_{\scriptscriptstyle \perp}$
describes the divergence  of the spatial correlation
length, i.e., 
$\xi_{\scriptscriptstyle \perp} \propto \delta\rho^{-\nu_{\scriptscriptstyle \perp}}$.
Note that the universal scaling functions 
depend now on the particular
choice of the boundary conditions,
the system shape etc.~\cite{PRIVMAN_3}.
Throughout this work we use in all dimensions hyper cubic lattices
with periodic boundary conditions (pbc).
However, the universal scaling functions 
[Eqs.\,(\ref{eq:scal_ansatz_EqoS},\ref{eq:scal_ansatz_Fluc})]
are recovered
in the thermodynamic limit, e.g.
\begin{equation}
{\tilde R}_{\scriptscriptstyle \text {pbc}}(x,y,\infty)
\; = \; {\tilde R}(x,y)\, .
\label{eq:td_limit_uni_fkt}
\end{equation}

Additionally to the order parameter and its fluctuations
we consider the fourth-order cumulant~$Q$ which
is defined as (see for instance~\cite{BINDER_1})
\begin{equation}
Q \; = \;
1 \, - \, \frac{\langle \rho_{\scriptscriptstyle \text a}^4 \rangle}
{\, 3\,\langle \rho_{\scriptscriptstyle \text a}^2 \rangle^2\,} \, .
\label{eq:def_binder_cum_op}
\end{equation}
For non-vanishing order-parameter 
the cumulant tends to $Q=2/3$ in the thermodynamic limit.
In the case of a zero order parameter 
the cumulant vanishes if the 
order parameter is characterized by a Gaussian distribution
symmetrically distributed around zero.
The latter case is observed in equilibrium systems,
e.g.~the Ising model for $T>T_{\scriptscriptstyle \text c}$.
In the case of absorbing phase transitions the order parameter 
is non-negative per definition.
Thus the order parameter is characterized by a non-trivial
distribution close to criticality and the above scenario does
not apply.

Nevertheless one expects that the cumulant obeys the
scaling form 
\begin{equation}
Q(\delta\rho, h, L) 
 \sim  
{\tilde Q}_{\scriptscriptstyle \text {pbc}}
(a_{\scriptscriptstyle \rho} \delta \rho \; \lambda, 
a_{\scriptscriptstyle h} h \, \lambda^{\sigma} ,
a_{\scriptscriptstyle L} L \,\lambda^{-\nu_{\perp}}) .
\label{eq:cum_op_scal}
\end{equation}
Notice that no metric factor $a_{\scriptscriptstyle \text Q}$
is needed since the cumulant is already dimensionless.
Choosing $a_{\scriptscriptstyle L} L \,\lambda^{-\nu_{\perp}}=1$
we get for zero-field
\begin{eqnarray}
Q(0, 0, L)  & = &
\left . {Q} \vphantom{\tilde Q}
(\delta\rho, 0, L) \right |_{\scriptscriptstyle \delta\rho=0} \nonumber \\ 
& \sim  &
\left .
{\tilde Q}_{\scriptscriptstyle \text {pbc}}
(a_{\scriptscriptstyle \rho} \delta \rho \; 
(a_{\scriptscriptstyle L} L)^{-\nu_{\perp}},  0, 1) \right |_{\scriptscriptstyle
\delta\rho=0} \nonumber \\  & = &
{\tilde Q}_{\scriptscriptstyle \text {pbc}}(0, 0, 1) 
\label{eq:cum_op_inter}
\end{eqnarray}
which is obviously universal.
The universal value ${\tilde Q}_{\scriptscriptstyle \text {pbc}}(0, 0, 1)$
corresponds to an intersection point if one plots
$Q$ as a function of $\rho$ for various system sizes~$L$.
Thus it is possible to determine the 
critical value~$\rho_{\scriptscriptstyle \text {c}}$ 
from the common intersection point.
This cumulant intersection method is very useful and was applied
in numerous works~(see for instance \cite{BINDER_1} and references
therein).

As usually finite-size effects have to be taken into account 
if the correlation length 
is of the order of the system size.
A feature of these finite-size effects is that 
a given system may pass within the simulations 
from one phase to the other.
This behavior is caused by critical fluctuations, i.e.,
approaching the transition point the order parameter
vanishes whereas its fluctuations diverge.
But in contrast to common second order phase transitions
the situation is drastically different in
the case of absorbing phase transitions.
Approaching the transition point the correlation
length $\xi_{\scriptscriptstyle \perp}$
increases and as soon as $\xi_{\scriptscriptstyle \perp}$
is of the order of $L$ the system may pass to an absorbing state
and is trapped forever.
Additionally to the absorbing phase 
($\langle \rho_{\scriptscriptstyle \text a}^k 
\rangle_{}=0$ for $\delta\rho<0$) the 
steady state order parameter and its higher moments 
vanish ($\langle \rho_{\scriptscriptstyle \text a}^k 
\rangle_{\scriptscriptstyle L}=0$) even in a small
vicinity above the critical point.
Thus in the case of absorbing phase transitions it is
impossible to consider finite-size effects of steady state
quantities around the critical point.

\begin{figure}[t]
  \includegraphics[width=8.0cm,angle=0]{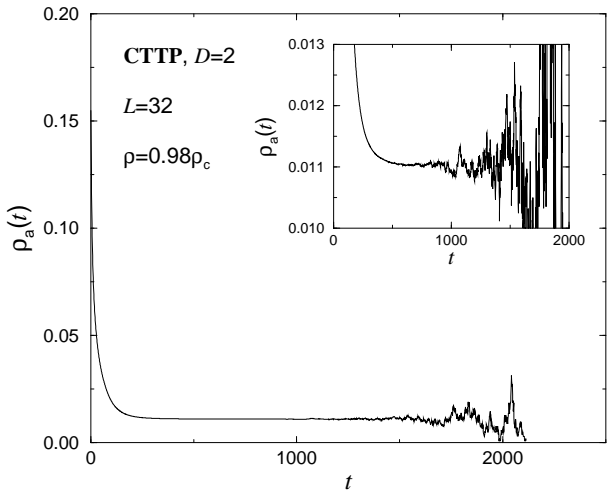}
  \caption{
    The decay of the order parameter close to criticality
    for the two-dimensional CTTP.
    After a so-called metastable regime the system
    passes to the absorbing phase.
    The inset displays that no clear saturation of the order
    parameter could be observed.
   }
  \label{fig:rho_a_t_ms}
\end{figure}

In order to bypass this problem it was 
suggested (see for instance~\cite{JENSEN_3})
to consider metastable (ms) or quasisteady state values 
of the order parameter and its
higher moments $\langle \rho_{\scriptscriptstyle \text a}^k 
\rangle_{\scriptscriptstyle L,\text {ms}}$,
respectively.
This is shown in Fig.\,\ref{fig:rho_a_t_ms} for the order
parameter of the two-dimensional CTTP close to the 
transition point.
After a short transient regime the system reaches a metastable
state where it can spend a certain time until 
it finally enters an absorbing state.
In the metastable phase the order parameter is expected
to fluctuate around a well defined average value 
$\langle \rho_{\scriptscriptstyle \text a}
\rangle_{\scriptscriptstyle L,\text {ms}}$
which is used for the finite-size scaling analysis.
This method was applied in 
previous works (e.g.~\cite{ROSSI_1,VESPIGNANI_4,JENSEN_3}
and the results sound mostly valid.

Nevertheless this method can be questioned.
First, there is no well defined average
value of the order parameter in the metastable
regime.
This can be seen in the inset of Fig.\,\ref{fig:rho_a_t_ms}
where we scrutinize the data.
No clear saturation of the order parameter can be
observed.
Second the method is quite inefficient.
In the case of the data presented in Fig.\,\ref{fig:rho_a_t_ms}
we used $5\, 10^{7}$ different initial configurations
for $L=32$ to get a sufficiently averaged estimate of the
order parameter.
Although roughly $10^{11}$ lattice updates 
($t_{\scriptscriptstyle \text {max}}\approx 2000$) 
were performed no clear saturation could be
observed.
Thus reliable data for larger lattice sizes 
($L=64,128,256,...$), which are required for 
an appropriate finite-size scaling analysis,
can not be obtained within moderate computer times.
Third and finally no rigorous prove exists that 
the metastable order parameter moments 
$\langle \rho_{\scriptscriptstyle \text a}^k 
\rangle_{\scriptscriptstyle L,\text {ms}}$
scale in the same way as the corresponding steady 
state order parameter moments.

\begin{figure}[t]
  \includegraphics[width=8.0cm,angle=0]{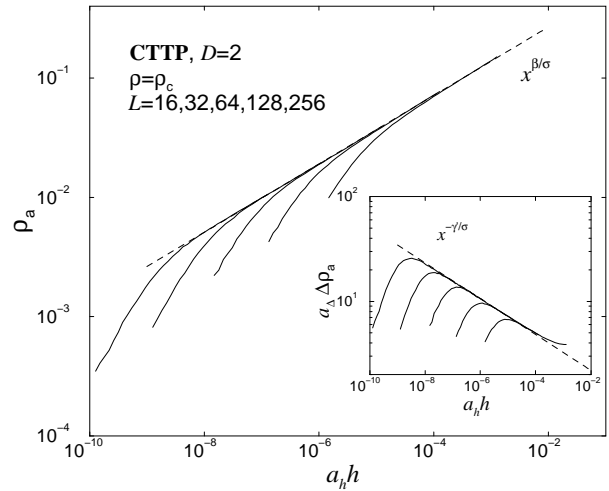}
  \caption{
    The density of active sites $\rho_{\text a}$ as a function
    of the external field for various system sizes~$L$.    
    The inset displays the corresponding order parameter
    fluctuations.
    The dashed-lines correspond to the asymptotic behavior
    of the infinite systems [Eqs.(\protect\ref{eq:metric_factors_a_h},
    \protect\ref{eq:metric_factors_a_Delta})].
   }
  \label{fig:rho_a_field_02}
\end{figure}

In contrast to the consideration of metastable phases
we choose a different method in order to study 
finite-size effects in the steady state. 
In our simulations we
measure the order parameter at the critical 
density ($\delta\rho=0$) as a function of the conjugated
field~$h$ for various system sizes.
Due to the external field the system can not be trapped forever 
in the absorbing phase.
Therefore steady state quantities are available 
for all densities.
In Fig.\,\ref{fig:rho_a_field_02} we present the order 
parameter and its fluctuations for the two-dimensional CTTP model.
The finite-size effects, i.e.,~the deviations from the behavior of
the "infinite" system 
($L\gg \xi_{\scriptscriptstyle \perp}$)
can be clearly seen.
Note that the data for $L=32$ are averaged 
over $3 \, 10^7$ lattice updates and we obtain smooth
curves of the order parameter, its fluctuations as well as 
of the cumulant.
Thus the numerical effort of this method is
significantly smaller
than the above discussed analysis of the metastable regime.

According to the above scaling laws 
[Eqs.\,(\ref{eq:scal_ansatz_EqoS_FSS},\ref{eq:cum_op_scal})]
the finite-size scaling forms are given by
\begin{eqnarray}
\label{eq:scal_EqoS_FSS}
\rho_{\scriptscriptstyle \text a}(0, h, L) 
& \sim & 
(a_{\scriptscriptstyle L} L)^{-\beta/ \nu_{\scriptscriptstyle \perp}}
{\tilde R}_{\scriptscriptstyle \text {pbc}}
(0, a_{\scriptscriptstyle h} h 
(a_{\scriptscriptstyle L} L)^{\sigma/ \nu_{\scriptscriptstyle \perp}}, 
1) , \nonumber \\
\label{eq:scal_Fluc_FSS}
a_{\scriptscriptstyle \Delta} 
\Delta \rho_{\scriptscriptstyle \text a}(0, h, L) 
& \sim & 
(a_{\scriptscriptstyle L} L)^{\gamma^{\prime}/ \nu_{\scriptscriptstyle \perp}} 
{\tilde D}_{\scriptscriptstyle \text {pbc}}
(0, 
a_{\scriptscriptstyle h} h 
(a_{\scriptscriptstyle L} L)^{\sigma/ \nu_{\scriptscriptstyle \perp}} ,
1) , \nonumber \\
\label{eq:scal_Cum_FSS}
Q(0, h, L) 
 &\sim  &
{\tilde Q}_{\scriptscriptstyle \text {pbc}}
(0, 
a_{\scriptscriptstyle h} h  
(a_{\scriptscriptstyle L} L)^{\sigma/ \nu_{\scriptscriptstyle \perp}} ,
1) .
\end{eqnarray}
For the sake of convenience we norm the universal
scaling function ${\tilde Q}_{\scriptscriptstyle {\text pbc}}$
by the condition
\begin{equation}
{\tilde Q}_{\scriptscriptstyle {\text pbc}}(0,1,1) \; = \;0.
\label{eq:metric_factors_a_L}
\end{equation}
Since the metric factor $a_{\scriptscriptstyle h}$ 
is known from previous simulations~\cite{LUEB_26}
(via Eq.\,(\ref{eq:metric_factors_a_h}))
the above condition can be used to determine 
the metric factor~$a_{\scriptscriptstyle L}$.
Taking into account that the correlation length scales
at criticality as 
\begin{equation}
a_{\scriptscriptstyle \perp} \, \xi_{\scriptscriptstyle \perp}
\; \sim \; (a_{\scriptscriptstyle h}\, h)^{-\nu_{\scriptscriptstyle \perp}/\sigma}
\label{eq:xi_h_at_rhoc}
\end{equation}
we find that Eq.\,(\ref{eq:metric_factors_a_L}) implies that
the universal function ${\tilde Q}_{\scriptscriptstyle {\text pbc}}$ 
is positive for $a_{\scriptscriptstyle L} L > 
a_{\scriptscriptstyle \perp} \, \xi_{\scriptscriptstyle \perp}$ 
and negative for 
$a_{\scriptscriptstyle L} L < 
a_{\scriptscriptstyle \perp} \, \xi_{\scriptscriptstyle \perp}$ .
Note that in the case of equilibrium phase transitions
Eq.\,(\ref{eq:metric_factors_a_L}) is useless since 
the cumulant is usually positive.

\begin{figure}[t]
  \includegraphics[width=8.0cm,angle=0]{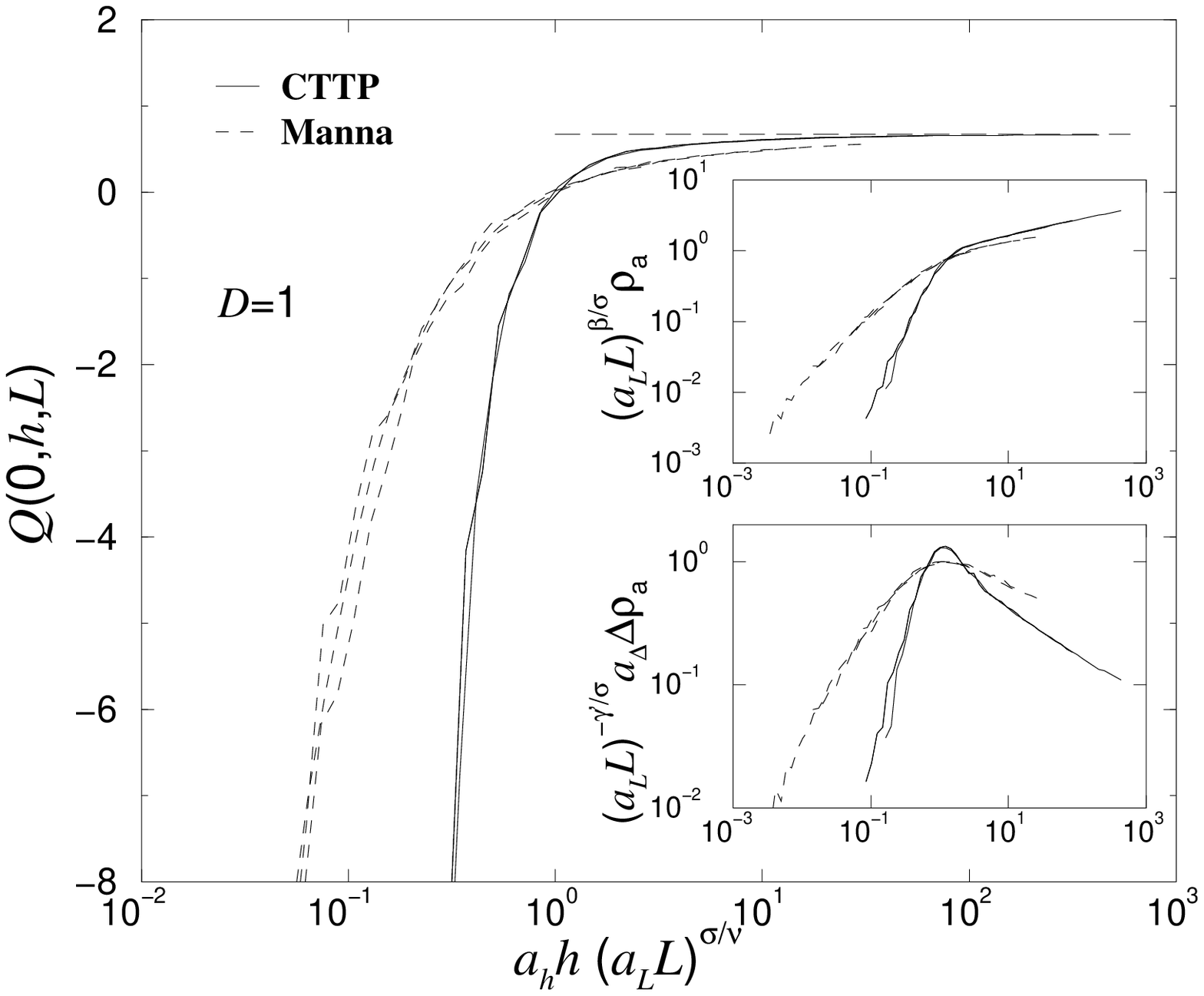}
  \caption{
    The universal finite-size scaling analysis of the 
    order parameter $\rho_{\scriptscriptstyle \text a}$,
    of the fluctuations $\Delta\rho_{\scriptscriptstyle \text a}$
    as well as of the cumulant $Q$ for the one dimensional
    CTTP and the one-dimensional Manna model.
    The long-dashed lines correspond to the power law behaviors
    of the infinite system [Eqs.(\protect\ref{eq:metric_factors_a_h},
    \protect\ref{eq:metric_factors_a_Delta})]
    and to the cumulant limit $2/3$, respectively.
    The data are obtained from simulations of system sizes
    $L\in \{2048,4096,8192 \}$ where up to $10^{10}$ lattice
    update steps are performed.
   }
  \label{fig:op_fss_field_1d}
\end{figure}

\begin{figure}[t]
  \includegraphics[width=8.0cm,angle=0]{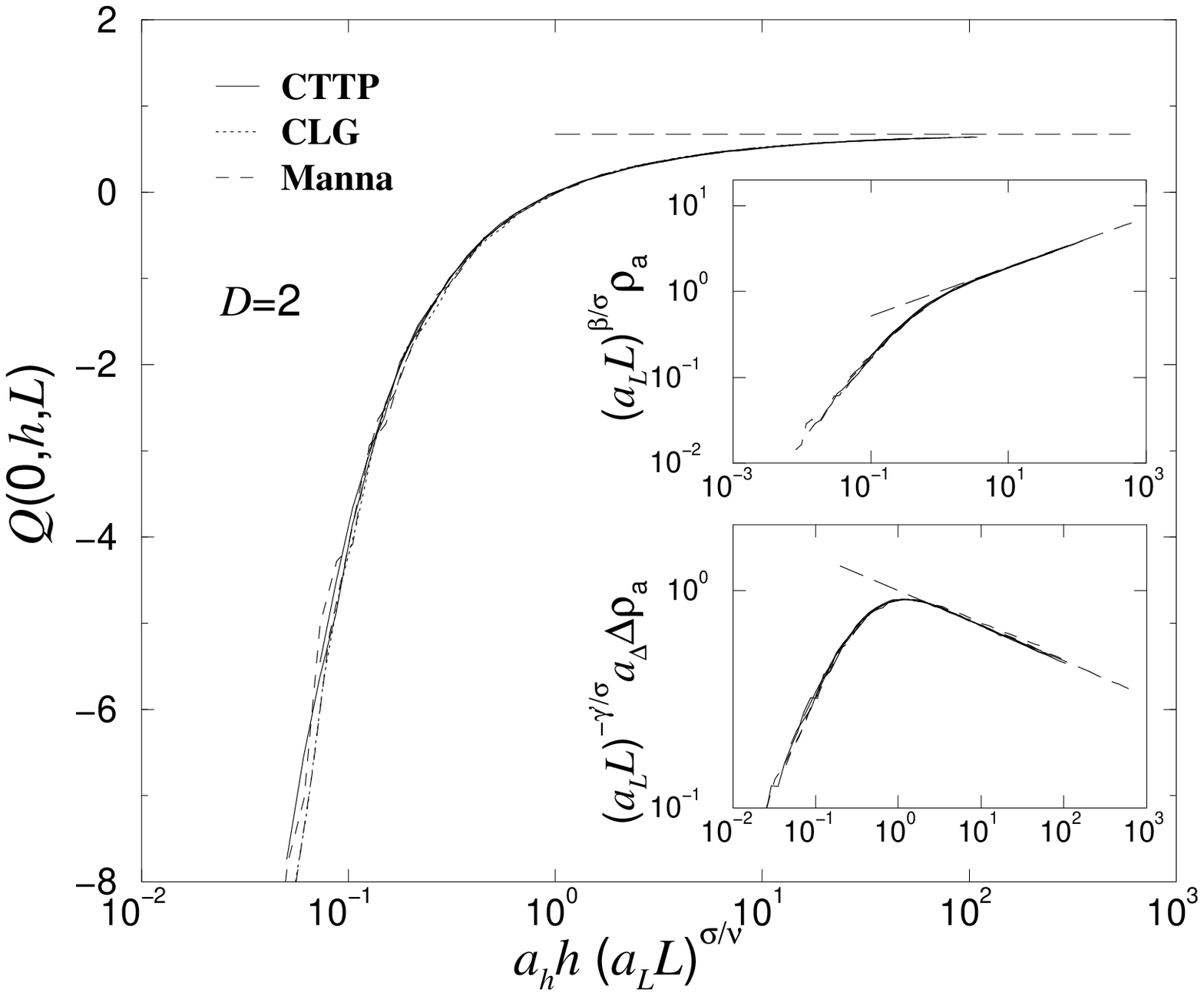}
  \caption{
    The universal finite-size scaling analysis of the 
    order parameter $\rho_{\scriptscriptstyle \text a}$,
    of the fluctuations $\Delta\rho_{\scriptscriptstyle \text a}$
    as well as of the cumulant $Q$ for the two dimensional
    models.
    The long-dashed lines correspond to the power law behaviors
    of the infinite system [Eqs.(\protect\ref{eq:metric_factors_a_h},
    \protect\ref{eq:metric_factors_a_Delta})]
    and to the cumulant limit $2/3$, respectively.
    The data are obtained from simulations of system sizes
    $L\in \{64,128,256\}$ where up to $5\,10^{7}$ lattice
    update steps are performed.\\[-1mm]
   }
  \label{fig:op_fss_field_2d}
%
  \includegraphics[width=8.0cm,angle=0]{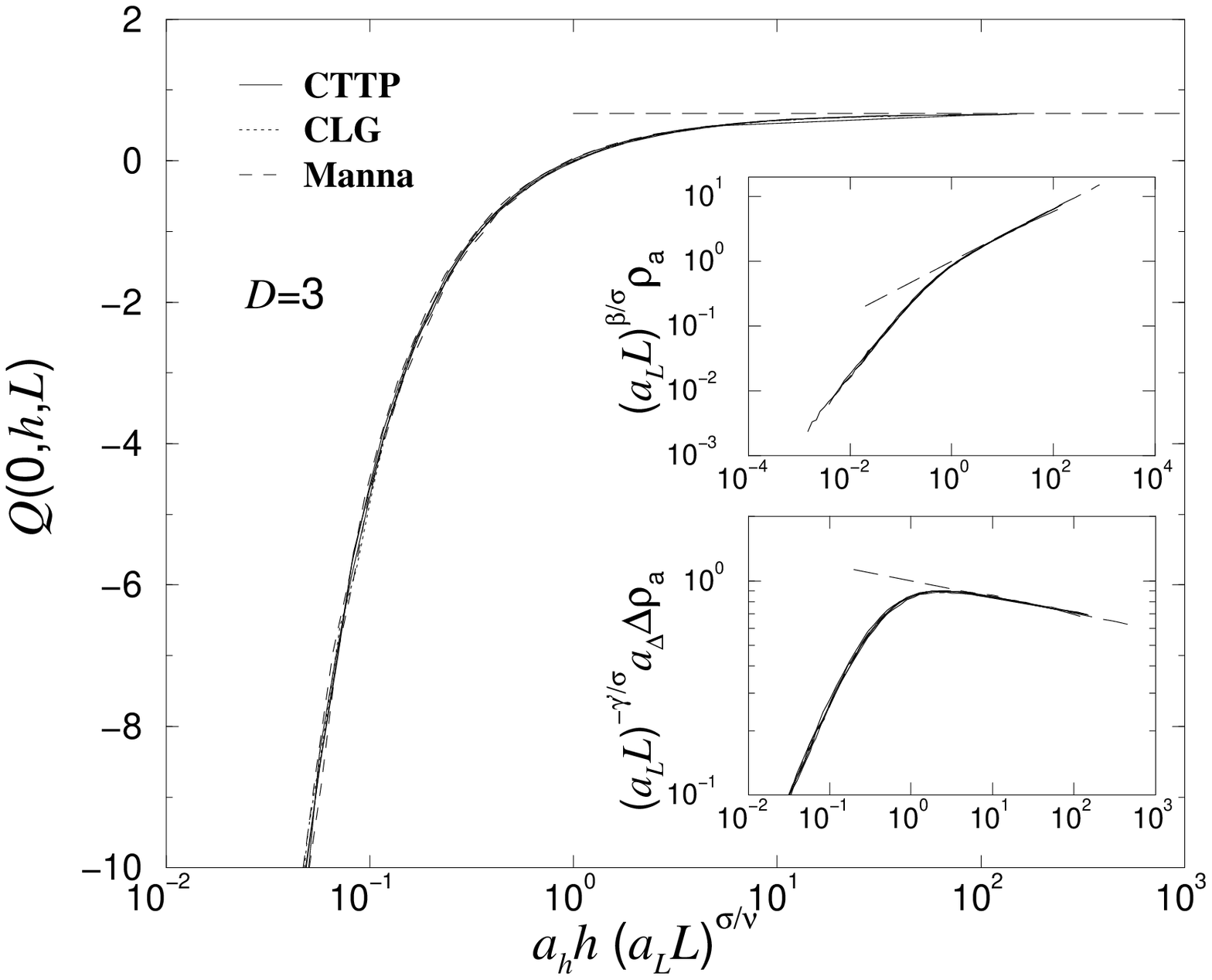}
  \caption{
    The universal finite-size scaling analysis of the 
    order parameter $\rho_{\scriptscriptstyle \text a}$,
    of the fluctuations $\Delta\rho_{\scriptscriptstyle \text a}$
    as well as of the cumulant $Q$ for the three dimensional
    models.
    The long-dashed lines correspond to the power law behaviors
    of the infinite system [Eqs.(\protect\ref{eq:metric_factors_a_h},
    \protect\ref{eq:metric_factors_a_Delta})]
    and to the cumulant limit $2/3$, respectively.
    The data are obtained from simulations of system sizes
    $L\in \{16,32,64\}$ where up to $5\, 10^{7}$ lattice
    update steps are performed.
   }
  \label{fig:op_fss_field_3d}
\end{figure}

In Figs.\,\ref{fig:op_fss_field_1d}-\ref{fig:op_fss_field_3d} 
we present the universal finite-size scaling analysis 
of the CLG model, the Manna model and the CTTP for $D=1,2,3$.
Since the exponents $\beta$ and $\sigma$ 
were already determined in previous works~\cite{LUEB_22,LUEB_24,LUEB_26}
we just vary the correlation length exponent 
$\nu_{\scriptscriptstyle \perp}$ in order to produce 
data collapses.
The value of the non-universal metric factor $a_{\scriptscriptstyle L}$
is determined via Eq.\,(\ref{eq:metric_factors_a_L}).
We observe good data collapses 
for $\nu_{\scriptscriptstyle \perp}=0.799\pm 0.014$  for $D=2$
and $\nu_{\scriptscriptstyle \perp}=0.593\pm 0.013$  for $D=3$,
respectively.
The data collapses are quite sensitive for variations of the
exponents.
Thus the quality of the corresponding data collapses
are used to estimate the error-bars.
The values of the exponents as well as of the non-universal
metric factors are listed in Table\,\ref{table:exponents}
and Table\,\ref{table:metric_factors}.

In the case of the one dimensional Manna model and the 
one dimensional CTTP we observe the expected splitting of the
universality class. 
The correlations length exponent $\nu_{\scriptscriptstyle \perp}$,
the field exponent $\sigma$, as well as the 
scaling functions differ clearly 
(see Table\,\ref{table:exponents} and Fig.\,\ref{fig:op_fss_field_1d}).
Furthermore the value of the Manna model 
$\nu_{\scriptscriptstyle \perp}=1.347\pm 0.091$ 
differs clearly from $\nu_{\scriptscriptstyle \perp}=1.80\pm 0.01$
obtained in a previous work including a finite-size scaling
analysis of metastable states~\cite{DICKMAN_2}.

Despite this splitting of universality we 
observe that the fourth-order cumulant tends 
for all models in all dimensions
to infinity if one approaches the transition
point, i.e.,
\begin{equation}
{\tilde Q}_{\scriptscriptstyle {\text {pbc}}}
(0,x,1) \; \to \; -\infty
\quad {\text {for}}
\quad
x\; \to \; 0 \,.
\label{eq:uni_value_Q_APT}
\end{equation}
This behavior is caused by the vanishing 
steady state fluctuations [see Eq.\,(\ref{eq:def_binder_cum_op})].
Thus we assume that the divergent fourth-order cumulant
is a characteristic feature of all absorbing 
phase transitions, independently of the considered lattice
structure as well as of the particular considered
universality class.
Preliminary simulations for directed percolation
support this conjecture and will be published elsewhere.

\section{Dynamical scaling behavior}

\subsection{Homogeneous particle source}

\begin{figure}[b]
  \includegraphics[width=8.0cm,angle=0]{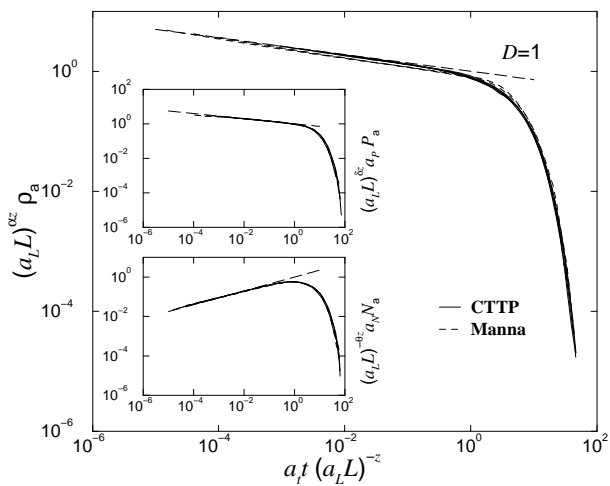}
  \caption{
    The dynamical scaling analysis for the one-dimensional 
    CTTP and the Manna model.
    The long-dashed lines correspond to the power law behaviors
    of the infinite systems [Eqs.\,(\protect\ref{eq:rho_a_time_crit},
    \protect\ref{eq:P_sur_at_crit},
    \protect\ref{eq:Na_at_crit})].
    System sizes from $L=512$ up to $L=8192$ are considered and 
    the data are averaged over at least $10^{5}$ different initial
    natural configurations (see text).
   }
  \label{fig:spread_uni_1d_01}
\end{figure}

In the following we investigate the dynamical
scaling behavior in the vicinity of the
absorbing phase transition.
First we consider how the order
parameter $\rho_{\scriptscriptstyle \text a}$ 
decays, starting the simulations
from a random distribution of particles
(so-called homogeneous particle source).
Above the transition point the density of active sites 
decreases in time and tends to the steady state
value (despite of finite-size 
effects as discussed above).
Below the transition point the density of active sites
decreases exponentially to zero.
At the critical point the
order parameter decays algebraically according to
\begin{equation}
\rho_{\scriptscriptstyle \text a} \; \sim  \; (a_{\scriptscriptstyle t} t)^{-\alpha}
\label{eq:rho_a_time_crit}
\end{equation}
where $a_{\scriptscriptstyle t}$ denotes a corresponding
non-universal metric factor.
A finite system size limits this power-law behavior
and one expects that the order parameter obeys at criticality
the scaling ansatz
\begin{equation}
\rho_{\scriptscriptstyle \text a}(L,t) \; \sim  \; 
\lambda^{-\alpha \nu_{\scriptscriptstyle\parallel}} \; 
{\tilde R}^{\prime}_{\scriptscriptstyle \text {pbc}} 
(
a_{\scriptscriptstyle t}  t \,
\lambda^{- \nu_{\scriptscriptstyle\parallel}} , 
a_{\scriptscriptstyle L} L \,
\lambda^{- \nu_{\scriptscriptstyle \perp}} )
\label{eq:ord_dyn_scal_R_prime}
\end{equation}
where we have to distinguish the universal functions
${\tilde R}^{\prime}$ and ${\tilde R}$.
For the sake of simplicity we choose ${\tilde R}^{\prime}(1,\infty)=1$.
Setting $a_{\scriptscriptstyle L} L \lambda^{-\nu_{\scriptscriptstyle \perp}}=1$ one gets
the  finite-size scaling form 
\begin{equation}
\rho_{\scriptscriptstyle \text a}(L,t) \; \sim  \; L^{-\alpha z} \,
{\tilde R}^{\prime}_{\scriptscriptstyle \text {pbc}}
(a_{\scriptscriptstyle t}  t \, (a_{\scriptscriptstyle L} L)^{-z} , 1)
\label{eq:ord_dyn_scal_FSS_01}
\end{equation}
where $z=\nu_{\scriptscriptstyle \parallel}
/\nu_{\scriptscriptstyle \perp}$ denotes the
dynamical exponent as usual.
Finite-size effects have to be taken into account
for ${\cal O}(t)=t_{\scriptscriptstyle \text {FSS}}$
with 
\begin{equation}
t_{\scriptscriptstyle \text {FSS}} \; = \;
a_{\scriptscriptstyle t}^{-1}   \, (a_{\scriptscriptstyle L} L)^{z} \, .
\label{eq:def_t_FSS}
\end{equation}
For $t \ll t_{\scriptscriptstyle \text {FSS}}$ the scaling 
function obeys the power-law 
${\tilde R}^{\prime}_{\scriptscriptstyle \text {pbc}} 
(x,1) \sim x^{-\alpha}$, whereas
${\tilde R}^{\prime}_{\scriptscriptstyle \text {pbc}} (x,1)$
decays exponentially for $x \gg 1$, i.e., 
$t \gg t_{\scriptscriptstyle \text {FSS}} $.

\begin{figure}[b]
  \includegraphics[width=8.0cm,angle=0]{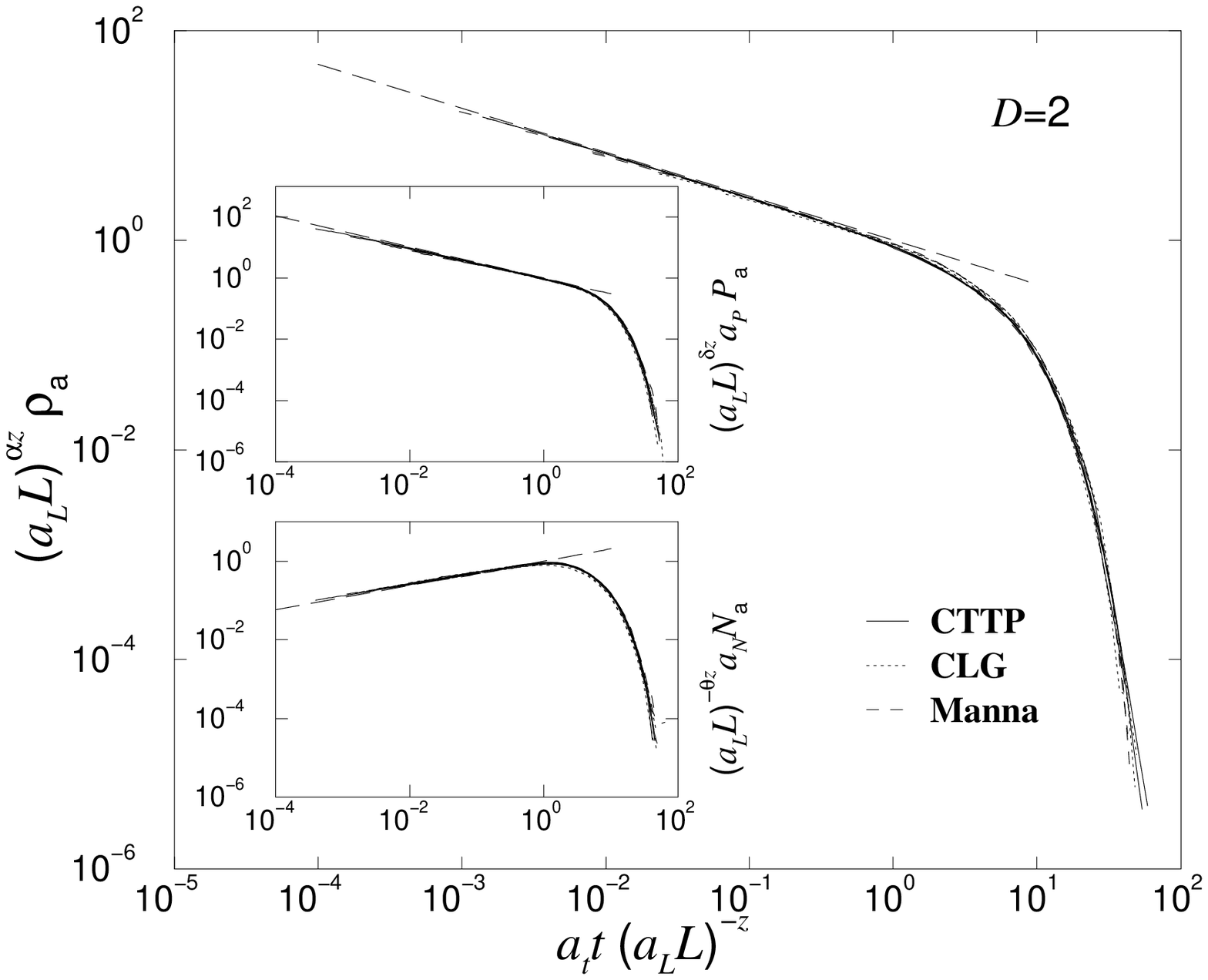}
  \caption{
    The dynamical scaling analysis for the two-dimensional models.
    The long-dashed lines correspond to the power law behaviors
    of the infinite systems [Eqs.\,(\protect\ref{eq:rho_a_time_crit},
    \protect\ref{eq:P_sur_at_crit},
    \protect\ref{eq:Na_at_crit})].
    System sizes from $L=64$ up to $L=512$ are considered and 
    the data are averaged over at least $2\, 10^{6}$ different initial
    natural configurations (see text).\\[-1mm]
   }
  \label{fig:spread_uni_2d_01}
  \includegraphics[width=8.0cm,angle=0]{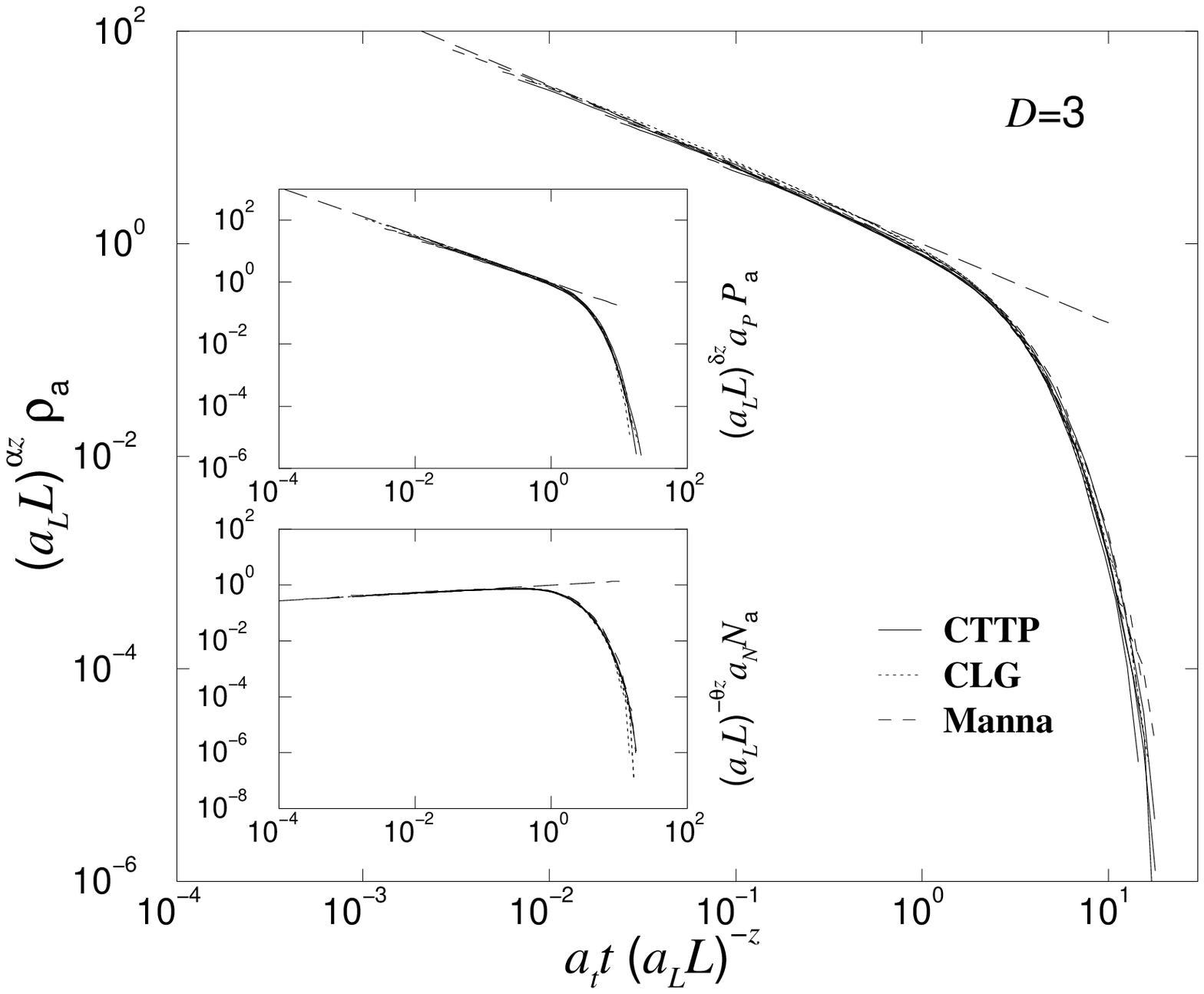}
  \caption{
    The dynamical scaling analysis for the three-dimensional models.
    The long-dashed lines correspond to the power law behaviors
    of the infinite systems [Eqs.\,(\protect\ref{eq:rho_a_time_crit},
    \protect\ref{eq:P_sur_at_crit},
    \protect\ref{eq:Na_at_crit})].
    System sizes from $L=16$ up to $L=128$ are considered and 
    the data are averaged over at least $5\, 10^{6}$ different initial
    natural configurations (see text).
   }
  \label{fig:spread_uni_3d_01}
\end{figure}

According to the above scaling form 
[Eq.\,(\ref{eq:ord_dyn_scal_FSS_01})] we plot
in the Figs.\,\ref{fig:spread_uni_1d_01}-\ref{fig:spread_uni_3d_01}
the rescaled order parameter as a function of the 
rescaled time.
We observe good data collapses for $\alpha=0.419\pm0.015$, 
$z=1.533\pm0.024$ for $D=2$ and 
$\alpha=0.745\pm0.017$, 
$z=1.823\pm0.023$ for $D=3$,
respectively.
These values are in agreement with those of previous
simulations~\cite{ROSSI_1}.
In the case of the one-dimensional models we observe
that the CTTP and the Manna model are both characterized
by $\alpha=0.141\pm0.024$ and $z=1.393\pm0.037$ 
but the corresponding scaling curves of both
models differ slightly.
It is possible that this indicates a universality 
splitting similar to the steady state scaling behavior.

\subsection{Localized particle source}

Additionally to a homogeneously distributed 
source of active sites one usually considers
the activity spreading generated from a single active
seed~\cite{GRASSBERGER_4}.
In this case it is customary to examine the survival probability 
$P_{\scriptscriptstyle \text a}(t)$ 
that the system is still active after $t$~update steps.
Furthermore one investigates how the number 
of active sites $N_{\scriptscriptstyle \text a}(t)$ increases 
in time.
At criticality the survival probability as well as the
average number of active sites are expected to 
scale as
\begin{eqnarray}
\label{eq:P_sur_at_crit_scal}
a_{\scriptscriptstyle P} \, 
P_{\scriptscriptstyle \text a}(L,t) & \sim &
\lambda^{-\delta \nu_{\scriptscriptstyle \parallel}} \,
{\tilde P}_{\scriptscriptstyle \text {pbc}}
(
a_{\scriptscriptstyle t}  t \,
\lambda^{- \nu_{\scriptscriptstyle\parallel}} , 
a_{\scriptscriptstyle L} L \,
\lambda^{- \nu_{\scriptscriptstyle \perp}} )
\, , \\
\label{eq:Na_at_crit_scal}
a_{\scriptscriptstyle N} \,    
N_{\scriptscriptstyle \text a}(L,t) & \sim  & 
\lambda^{\theta \nu_{\scriptscriptstyle \parallel}} \,
{\tilde N}_{\scriptscriptstyle \text {pbc}}
(
a_{\scriptscriptstyle t}  t \,
\lambda^{- \nu_{\scriptscriptstyle\parallel}} , 
a_{\scriptscriptstyle L} L \,
\lambda^{- \nu_{\scriptscriptstyle \perp}} )  \, .
\end{eqnarray}
The universal functions are normed by ${\tilde P}(1,\infty)=1$
and ${\tilde N}(1,\infty)=1$ and 
we find in the thermodynamic limit
\begin{eqnarray}
\label{eq:P_sur_at_crit}
a_{\scriptscriptstyle P} \,    
P_{\scriptscriptstyle \text a} & \sim  &
(a_{\scriptscriptstyle t}  t)^{-\delta} \, ,  \\
\label{eq:Na_at_crit}
a_{\scriptscriptstyle N} \, 
N_{\scriptscriptstyle \text a} & \sim  &
(a_{\scriptscriptstyle t}  t)^{\theta} \, .
\end{eqnarray}
The finite-size scaling forms are obtained
by setting $a_{\scriptscriptstyle L} L 
\lambda^{- \nu_{\scriptscriptstyle \perp}} =1$, yielding
\begin{eqnarray}
\label{eq:P_sur_at_crit_FSS_01}   
a_{\scriptscriptstyle P} \,    
P_{\scriptscriptstyle \text a}(L,t) & \sim  & L^{-\delta z} \,
{\tilde P}_{\scriptscriptstyle \text {pbc}}
(a_{\scriptscriptstyle t}  t \, (a_{\scriptscriptstyle L} L)^{-z} , 1) \, , \\
\label{eq:Na_at_crit_FSS_01}
a_{\scriptscriptstyle N} \,    
N_{\scriptscriptstyle \text a}(L,t) &\sim  & L^{\theta z} \,
{\tilde N}_{\scriptscriptstyle \text {pbc}}
(a_{\scriptscriptstyle t}  t \, (a_{\scriptscriptstyle L} L)^{-z} , 1)
\, .
\end{eqnarray}
Again the scaling functions ${\tilde P}_{\scriptscriptstyle \text {pbc}}$
and ${\tilde N}_{\scriptscriptstyle \text {pbc}}$
decay exponentially for $t \gg t_{\scriptscriptstyle \text {FSS}}$
whereas they exhibit an algebraic behavior
for $t \ll t_{\scriptscriptstyle \text {FSS}}$.

Since the absorbing state is non-trivial
one has to investigate the spreading activity at the 
so-called natural density (see for instance~\cite{JENSEN_3}).
For each considered model an absorbing state at 
$\rho_{\scriptscriptstyle \text c}$ is prepared and a particle
is moved to a randomly selected site in order to create one active seed.
We obtain for all dimensions convincing data collapses
which are shown in the insets of the 
Figs.\,\ref{fig:spread_uni_1d_01}-\ref{fig:spread_uni_3d_01}.
The values of the exponents agrees with those of previous 
works~\cite{ROSSI_1} and are listed together with the 
non-universal metric factors in Table\,\ref{table:exponents}
and Table\,\ref{table:metric_factors}.
In summary activity spreading from a localized seed
is characterized for all three models 
by the same universal scaling functions 
${\tilde P}$ and ${\tilde N}$.

\begin{table}[t]
\caption{The critical exponents of the considered
models below the upper critical dimension 
$D_{\scriptscriptstyle \text c}=4$.
The data of the exponents $\beta$ and $\sigma$ are 
obtained from~\protect\cite{LUEB_24}.
The finite-size scaling analysis yields the values of
$\nu_{\scriptscriptstyle \perp}$ and $\gamma^{\prime}$,
whereas the exponents $\alpha$, $\delta$, $\theta$, and $z$ 
are obtained from activity spreading (see text).
The values of $\beta^{\prime}$ and $\nu_{\scriptscriptstyle \perp}$ 
are determined via scaling laws.
In particular the values of $\nu_{\scriptscriptstyle \parallel}$ are 
in good agreement with those of direct measurements of the order 
parameter persistence distribution~\protect\cite{LUEB_21}. 
In the case of the one-dimensional models we observe a 
splitting of the universality class.}
\label{table:exponents}
\begin{tabular}{llll}
$$       &  $D=1$	& $D=2$	& $D=3$ \\  
\tableline 
$\alpha\quad$&  $0.141\pm0.024$	    & $0.419\pm0.015$   & $0.745\pm0.017$ \\ 
$\delta$     &  $0.170\pm0.025$     & $0.510\pm0.020$   & $0.765\pm0.025$ \\ 
$\theta$     &  $0.350\pm0.030$     & $0.310\pm0.030$   & $0.140\pm0.030$ \\
$z$ 	     &  $1.393\pm0.037$	    & $1.533\pm0.024$   & $1.823\pm0.023$ \\      
\tableline
$\beta$      &  $0.382\pm0.019$	& $0.639\pm0.009\quad$ & $0.840\pm0.012$ \\   
$\beta^{\prime}$  &  $0.319\pm0.052_{\scriptscriptstyle\text
{Manna}}$	& $0.624\pm0.029\quad$ & $0.827\pm0.034$ \\ 
$\sigma$     &  $2.710\pm0.040_{\scriptscriptstyle\text {Manna}}\quad$	& $2.229\pm0.032$ &
$2.069\pm0.043$\\    
$$           &  $1.770\pm0.058_{\scriptscriptstyle\text {CTTP}}\quad$	& $$ & $$ \\   
$\nu_{\scriptscriptstyle \perp}$     &  $1.347\pm0.091_{\scriptscriptstyle\text
{Manna}}$	& $0.799\pm0.014$ & $0.593\pm0.013$\\    
$$           &  $1.760\pm0.060_{\scriptscriptstyle\text {CTTP}}$	& $$ & $$ \\   
$\nu_{\scriptscriptstyle \parallel}$     &  $1.876\pm 0.135_{\scriptscriptstyle\text
{Manna}}$	& $1.225\pm0.029$ & $1.081\pm0.027$\\    
$$           &  $2.452\pm0.106_{\scriptscriptstyle\text {CTTP}}$	& $$ & $$ \\   
$\gamma^{\prime}$     &  $0.550\pm0.040_{\scriptscriptstyle\text {Manna}}$	&
$0.367\pm0.019$ & $0.152\pm0.017$\\    
$$           &  $0.670\pm0.040_{\scriptscriptstyle\text {CTTP}}$	& $$ & $$ 
\end{tabular}
\end{table}

The activity spreading of APT
with a conserved field is closely connected to 
avalanche processes in SOC systems~\cite{VESPIGNANI_4}.
In particular the Manna model is a paradigmatic
example of a class of SOC systems, the so-called
sandpile models.
The SOC version and the APT version of the Manna model
are characterized by the same (microscopic) dynamic
rules but the boundary conditions differ.
Closed boundary conditions lead to a globally
conserved particle density in the case of absorbing
phase transitions.
Whereas sandpile models are per definition driven 
dissipative systems where particles (sand-grains)
are injected into the system and dissipated through
open boundaries.
The self-organization of SOC systems corresponds to the
fact that they approach, without any external fine
tuning, the critical state 
($\rho(t) \to \rho_{\scriptscriptstyle \text c}$)
in the infinitesimally slow driving limit
(so-called separation of time scales, see~\cite{GRINSTEIN_1}).
In the critical state the external driving triggers
(scale invariant) avalanche-like relaxation events.
These avalanche processes are described by certain
critical exponents which can be derived from the spreading
exponents $\delta$, $\theta$, and $z$~\cite{VESPIGNANI_4,MUNOZ_3}.
In particular the avalanches are characterized by several
quantities (see for instance~\cite{BENHUR_1,LUEB_4}), e.g.~the
size~$s$ (number of elementary relaxation events),
the area~$a$ (number of distinct toppled sites),
the time~$t$ (number of parallel updates until 
the configuration is stable),
as well as the radius exponent~$r$ (radius of gyration).
In the critical steady state the corresponding
probability distributions decay algebraically 
\begin{equation}
P_{\scriptscriptstyle x} \; \propto \;
x^{-\tau_{\scriptscriptstyle x}} \, ,
\label{eq:soc_prob_tau_x}
\end{equation}
characterized by the avalanche exponents 
$\tau_{\scriptscriptstyle x}$ with
$x\in \{s,a,t,r \}$.
Assuming that the size, area, etc.~scale as
a power of each other,
\begin{equation}
x \; \propto \;
{x^{\prime}}^{\gamma_{\scriptscriptstyle xx^{\prime}}} \,
\label{eq:soc_gamma_xx}
\end{equation}
one obtains the scaling relations 
\begin{equation}
{\gamma_{\scriptscriptstyle xx^{\prime}}}
\; = \;
\frac{\,\tau_{\scriptscriptstyle x^\prime}-1\,}{\,\tau_{\scriptscriptstyle x}-1\,} \; .
\label{eq:soc_gamma_xx_taux_taux}
\end{equation}
The exponent~$\gamma_{\scriptscriptstyle tr}$ equals 
the dynamical exponent~$z$,
the exponent~$\gamma_{\scriptscriptstyle ar}$ corresponds to
the fractal dimension of the avalanches and the 
exponent~$\gamma_{\scriptscriptstyle sa}$ indicates
whether multiple toppling events are relevant 
($\gamma_{\scriptscriptstyle sa}>1$) 
or irrelevant ($\gamma_{\scriptscriptstyle sa}=1$).

\begin{figure}[t]
  \includegraphics[width=8.0cm,angle=0]{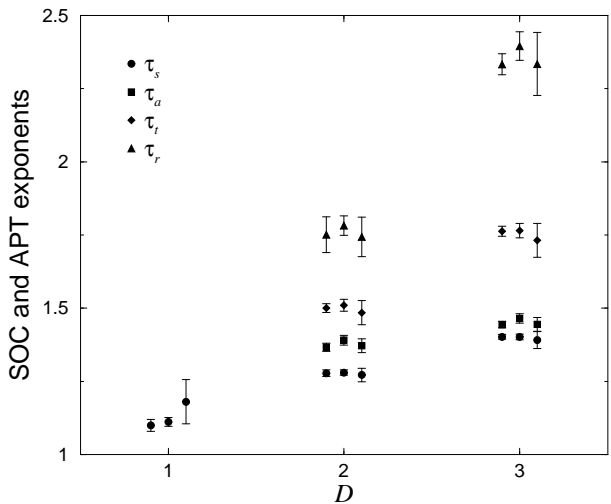}
  \caption{
    The avalanche exponents of the Manna model
    in various dimensions.
    In order to avoid overlaps the exponents are slightly shifted.
    The avalanche exponents of the SOC version of the Manna model
    (left) are obtained from~\protect\cite{DICKMAN_6} for $D=1$
    and from~\protect\cite{LUEB_PHD,LUEB_9} for $D=2,3$.
    Using the Eqs.\,(\protect\ref{eq:soc_tau_t}-\protect\ref{eq:soc_tau_s})
    we obtained the avalanche exponents (middle) from the spreading
    exponents $\delta$, $\theta$, and $z$.
    Using certain hyperscaling relations 
    it is possible to express the avalanche exponents (right)
    in terms of the exponents of the continuous
    absorbing phase transition, 
    see Eqs.\,(\protect\ref{eq:soc_exp_apt_exp_tau_r}-\protect\ref{eq:soc_exp_apt_exp_tau_s}).
   }
  \label{fig:soc_scal_01}
\end{figure}

These avalanche exponents are connected to the
spreading exponents $\delta$, $\theta$, and 
$z$ (see for instance ~\cite{MUNOZ_3}).
First the survival 
probability~$P_{\scriptscriptstyle \text a}(t)$ is simply
given by the integrated avalanche duration
\begin{equation}
P_{\scriptscriptstyle \text a}(t) \; = \;
\sum_{t^{\prime}=t}^{\infty} \, P_{\scriptscriptstyle t}(t^{\prime})
\label{eq:soc_prob_a_prob_t}
\end{equation}
yielding
\begin{equation}
\tau_{\scriptscriptstyle t} \; = \;
1 \, + \, \delta \, .
\label{eq:soc_tau_t}
\end{equation}
Since $\gamma_{\scriptscriptstyle tr}=z$ the radius exponent
is given by
\begin{equation}
\tau_{\scriptscriptstyle r} \; = \;
1 \, + \, z\, \delta \, .
\label{eq:soc_tau_r}
\end{equation}
Taking into account that the avalanches of the Manna model
are compact ($\gamma_{\scriptscriptstyle ar}=D$) 
below the upper critical dimension 
$D_{\scriptscriptstyle \text c}=4$~\cite{LUEB_4,LUEB_PHD}
we find 
\begin{equation}
\tau_{\scriptscriptstyle a} \; = \;
1 \, + \, \frac{\,z\, \delta\,}{D} \, .
\label{eq:soc_tau_a}
\end{equation}
Finally the number of topplings~$s_{\scriptscriptstyle t}$ 
for an avalanche that is active at time~$t$ equals the 
integrated numbers of active sites, i.e.,
\begin{equation}
s_{\scriptscriptstyle t} \, P_{\scriptscriptstyle \text a}(t)\; = \;
\sum_{{t^\prime} =0}^{t} \, N_{\scriptscriptstyle \text a}(t^{\prime})
\label{eq:soc_size_Na}
\end{equation}
leading to~\cite{MUNOZ_3}
\begin{equation}
\tau_{\scriptscriptstyle s} \; = \;
1 \, + \, \frac{\delta }{\, 1 + \theta + \delta \,} \, .
\label{eq:soc_tau_s}
\end{equation}
Thus the avalanche exponents of the Manna model are 
naturally related to the spreading exponents of the absorbing
phase transition.
In Fig.\,\ref{fig:soc_scal_01} we compare our results with
those of SOC simulations of the Manna model.
The data show that the above scaling relations 
[Eqs.\,(\ref{eq:soc_tau_t}-\ref{eq:soc_tau_s})] are fulfilled.

\section{Discussion}

\begin{figure}[b]
  \includegraphics[width=8.0cm,angle=0]{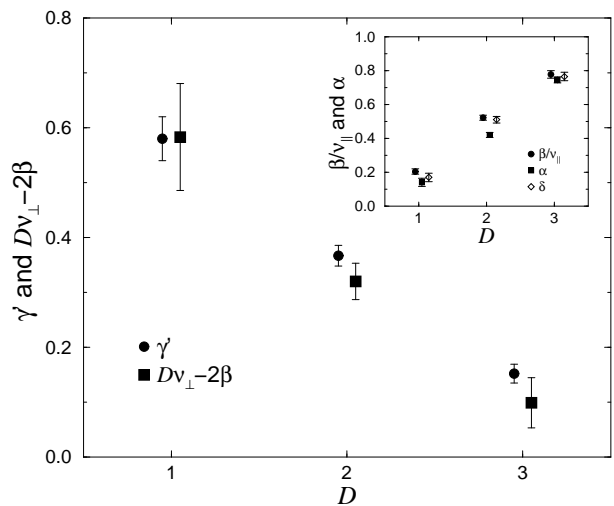}
  \caption{
    Test of the scaling relations 
    $\gamma^{\prime}= D \nu_{\scriptscriptstyle \perp} - 2 \beta$,
    $\alpha = \beta/\nu_{\scriptscriptstyle \parallel}$
    (inset), as well as 
    $\delta = \beta/\nu_{\scriptscriptstyle \parallel}$ (inset).
   }
  \label{fig:hyper_scal_01}
\end{figure}

In this section we check several scaling relations.
Due to the pathological behavior of the one-dimensional
CTTP we use the corresponding values of the Manna model
for our consideration.
At the beginning we check the scaling relation
\begin{equation}
\gamma^{\prime} \; = \;
D \, \nu_{\scriptscriptstyle \perp} \, - 2 \, \beta  \,
\label{eq:gamp_D_nu_beta}
\end{equation}
which can be easily derived from the scaling form of the
order parameter histogram (see for instance~\cite{JENSEN_3}).
The corresponding data are plotted in Fig.\,\ref{fig:hyper_scal_01}.
As can be seen the above scaling relation is fulfilled 
within the error-bars.

\begin{figure}[b]
  \includegraphics[width=8.0cm,angle=0]{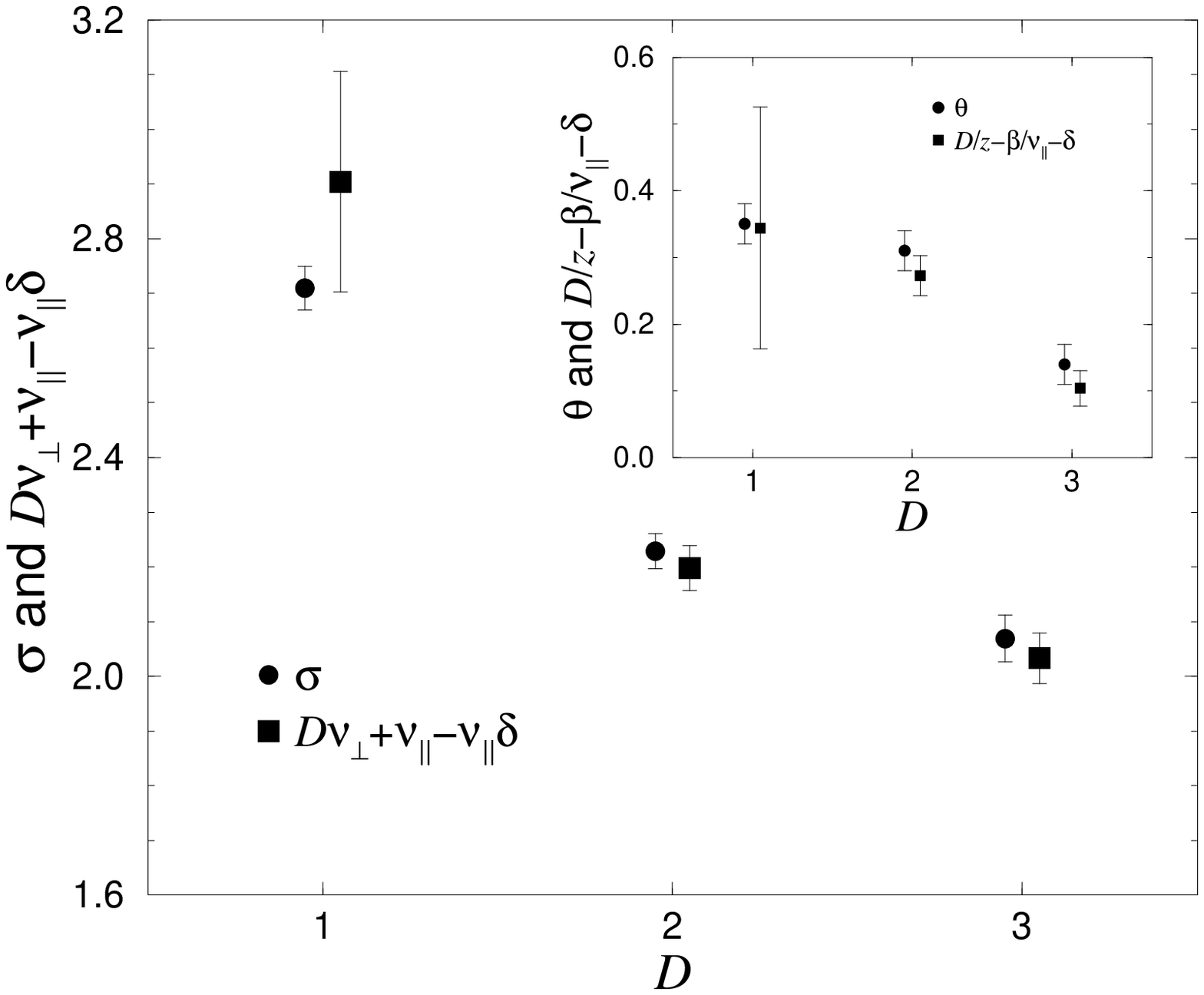}
  \caption{
    Test of the hyperscaling relations
    $\sigma=D \nu_{\scriptscriptstyle \perp}+\nu_{\scriptscriptstyle \parallel}
    - \nu_{\scriptscriptstyle \parallel} \delta$.
    and $\theta = D / z 
    - \beta / \nu_{\scriptscriptstyle \parallel} 
    - \delta$ (see inset). }
  \label{fig:hyper_scal_02}
\end{figure}

Taking into consideration that a weak external field
may trigger spreading events one finds that the
field exponent is given by~\cite{HINRICHSEN_1}
\begin{equation}
\sigma \; = \;
D \, \nu_{\scriptscriptstyle \perp} \, + \,
\nu_{\scriptscriptstyle \parallel} \, -  \, \nu_{\scriptscriptstyle \parallel} \,\delta  \, .
\label{eq:sigma_D_nu_perp_nu_par_nu_par_delta}
\end{equation}
In Fig.\,\ref{fig:hyper_scal_02} we check this scaling law.
As can be seen it is fulfilled 
within the error-bars.

Next we consider the hyperscaling relation
\begin{equation}
\theta \; = \;
\frac{\,D\,}{z}  \, - \,
\frac{\, \beta \,}{\nu_{\scriptscriptstyle \parallel}}
\, - \, \delta.
\label{eq:theta_D_z_delta_beta_nu_par}
\end{equation}
This scaling relation can be derived if one assumes that the 
steady state scaling forms and the dynamical scaling forms
can be combined to
\begin{eqnarray}
\label{eq:scal_ansatz_Na_rho}
& & N_{\scriptscriptstyle \text a}(\delta\rho, h, L,t,\rho_{\scriptscriptstyle \text a,0})
  =  L^D \; \rho_{\scriptscriptstyle \text a}
(\delta\rho, h, L,t,\rho_{\scriptscriptstyle \text a,0} ) \\
& \sim &  L^{D} \, \lambda^{-\beta} {\tilde R}_{\scriptscriptstyle \text {pbc}}
(a_{\scriptscriptstyle \rho}  
\delta \rho  \lambda, a_{\scriptscriptstyle h} h 
\lambda^{\sigma}, \nonumber \\
& & a_{\scriptscriptstyle L} L \lambda^{-\nu_{\scriptscriptstyle \perp}}, 
a_{\scriptscriptstyle t} t \lambda^{-\nu_{\scriptscriptstyle \parallel}},
a_{\scriptscriptstyle 0} 
\rho_{\scriptscriptstyle \text a,0} 
\lambda^{D \nu_{\scriptscriptstyle \perp}-\nu_{\scriptscriptstyle \parallel} \delta}) \, . 
\nonumber 
\end{eqnarray}
Here the initial density of active sites $\rho_{\scriptscriptstyle \text a,0}$
appears as an additional scaling field 
(see for instance~\cite{HINRICHSEN_1} and references therein)
and the scaling function behaves asymptotically as 
\begin{equation}
{\tilde R}_{\scriptscriptstyle \text {pbc}}(0,0,\infty,1,x)
\; \sim \;
\left \{
\begin{array}{ll}
x & \text{for} \quad x \ll 1 \\
\text{const} & \text{for} \quad x \gg 1  \, .
\end{array}
\right .
\label{eq:asymptotic_R_rho0}
\end{equation}
Starting at criticality from a low density of active 
sites (e.g.~several seeds)
the number of active sites increases as 
$N_{\scriptscriptstyle \text a} \propto t^{\theta}$
until it reaches a maximum 
and crosses over to the expected asymptotic decay 
$\rho_{\scriptscriptstyle \text a} \propto t^{-\alpha}$.
The crossover time is determined by 
\begin{equation}
{\cal O}( a_{\scriptscriptstyle 0} \rho_{\scriptscriptstyle \text a,0} \,
( a_{\scriptscriptstyle t} t_{\scriptscriptstyle \text{co}})^{D/z-\delta} )
\; = \; 1
\label{eq:cross_over_initial_slip}
\end{equation}
which corresponds to a merging of the 
survived (and former separated) clusters 
of activity~\cite{HINRICHSEN_1}.
The scaling of the crossover time
explains the choice of the scaling exponents in 
Eq.\,(\ref{eq:scal_ansatz_Na_rho}).
Setting $a_{\scriptscriptstyle t} t \lambda^{-\nu_{\scriptscriptstyle \parallel}}=1$
we find at criticality
\begin{eqnarray}
\label{eq:scal_ansatz_Na_rho_crit}
& & N_{\scriptscriptstyle \text a}(0, 0, L,t,\rho_{\scriptscriptstyle \text a,0}) 
\sim  L^D \; 
(a_{\scriptscriptstyle t} t)^{-\beta/ \nu_{\scriptscriptstyle \parallel}} \,  \\
& & {\tilde R}_{\scriptscriptstyle \text {pbc}}
(0,0,
a_{\scriptscriptstyle L} L (a_{\scriptscriptstyle t} t)^{-1/z},1,
a_{\scriptscriptstyle 0} \rho_{\scriptscriptstyle \text a,0} 
(a_{\scriptscriptstyle t} t)^{D / z - \delta} ) \, . 
\nonumber 
\end{eqnarray}
The full crossover can be observed if the particular 
value of the initial density of active sites 
$\rho_{\scriptscriptstyle {\text a},0}$ leads to 
$1 \ll t_{\scriptscriptstyle \text {co}}
\ll t_{\scriptscriptstyle \text {FSS}}$~(as it was for instance 
observed for directed percolation~\cite{HINRICHSEN_1}).
Therefore, the  first regime [Eq.\,(\ref{eq:Na_at_crit})] 
is obtained for 
$1 \ll t \ll t_{\scriptscriptstyle \text {co}}$,
the second regime [Eq.\,(\ref{eq:rho_a_time_crit})] for 
$t_{\scriptscriptstyle \text {co}} \ll t \ll t_{\scriptscriptstyle \text {FSS}}$
and finite-size effects take place in the third regime
for $t > t_{\scriptscriptstyle \text {FSS}}$.

In the case that one starts the simulations with a 
single seed $\rho_{\scriptscriptstyle \text a,0}=L^{-D}$
we get 
$t_{\scriptscriptstyle \text {co}} \propto L^{z D /(D-z \delta)}$.
Taking into account that $z < z D /(D-z \delta)$ 
we find 
that $t_{\scriptscriptstyle \text {FSS}} < t_{\scriptscriptstyle \text {co}}$, 
i.e., finite size effects take place before the algebraic
decay of active particles starts [Eq.\,(\ref{eq:rho_a_time_crit})]
and the second scaling regime does not occur.
Furthermore we can use for $t \ll t_{\scriptscriptstyle \text {FSS}}$ 
the approximation 
\begin{eqnarray}
\label{eq:Na_rho_crit_single_seed}
& & {\tilde R}_{\scriptscriptstyle \text {pbc}}
(0,0,
a_{\scriptscriptstyle L} L (a_{\scriptscriptstyle t} t)^{-1/z},1,
a_{\scriptscriptstyle 0} L^{-D} 
(a_{\scriptscriptstyle t} t)^{D / z - \delta} ) 
\nonumber \\
& \approx & {\tilde R}_{\scriptscriptstyle \text {pbc}}
(0,0, \infty,1,
a_{\scriptscriptstyle 0} \rho_{\scriptscriptstyle \text a,0} 
(a_{\scriptscriptstyle t} t)^{D / z - \delta} )  \nonumber \\
& \sim & a_{\scriptscriptstyle 0} \, L^{-D} \,
(a_{\scriptscriptstyle t} t)^{D / z - \delta}
\end{eqnarray}
and Eq.\,(\ref{eq:scal_ansatz_Na_rho_crit}) reads now
\begin{equation}
N_{\scriptscriptstyle \text a}
\; \sim \;
a_{\scriptscriptstyle 0} \,
(a_{\scriptscriptstyle t} t)^{-\beta/ \nu_{\scriptscriptstyle \parallel} 
+ D/z - \delta} \, . 
\label{eq:Na_early_time}
\end{equation}
Comparing this result with Eq.\,(\ref{eq:Na_at_crit}) 
we obtain the hyperscaling
relation Eq.\,(\ref{eq:theta_D_z_delta_beta_nu_par}) 
as well as $a_{\scriptscriptstyle 0} = 1 / a_{\scriptscriptstyle N}$.
In the inset of Fig.\,\ref{fig:hyper_scal_02} we display the
data of the corresponding exponents.
The hyperscaling relation Eq.\,(\ref{eq:theta_D_z_delta_beta_nu_par}) 
is fulfilled within the error-bars.

The situation is completely different if one 
starts the simulations with a homogeneous
particle source.
For instance a random distribution of particles leads
for the two-dimensional CTTP to an initial density
$\rho_{\scriptscriptstyle \text a,0}\approx 0.1703$.
In that case the crossover time
\begin{equation}
t_{\scriptscriptstyle \text{co}} \; =  \;
\frac{\, 1 \,}{a_{\scriptscriptstyle t}} \,
\left ( \frac{\,1\,}{a_{\scriptscriptstyle N}} \, \rho_{\scriptscriptstyle \text a,0}
\right )^{-\frac{1}{D/z-\delta}} 
\; \approx \; 1.14
\label{eq:cross_homo_source}
\end{equation}
is too small and the short time scaling regime 
($N_{\scriptscriptstyle \text a} \propto t^\theta$ for 
$1 \ll t \ll t_{\scriptscriptstyle \text {co}}$)
can not be observed.
On the other hand the scaling form 
Eq.\,(\ref{eq:scal_ansatz_Na_rho})
yields for the second regime
($t_{\scriptscriptstyle \text{co}} \ll t \ll t_{\scriptscriptstyle \text{FSS}}$)
\begin{eqnarray}
\label{eq:scal_rho_alpha}
& & \rho_{\scriptscriptstyle \text a} 
(\delta\rho, h, L,t,\rho_{\scriptscriptstyle \text a,0} )    \\
& \sim & (a_{\scriptscriptstyle t} t)^{-\beta/ \nu_{\scriptscriptstyle \parallel}}
{\tilde R}_{\scriptscriptstyle \text {pbc}} 
(0, 0,  
a_{\scriptscriptstyle L} L (a_{\scriptscriptstyle t} t)^{-1/z},1,
a_{\scriptscriptstyle 0} \rho_{\scriptscriptstyle \text a,0} 
t^{D / z - \delta} ) \nonumber \\
& \approx & 
(a_{\scriptscriptstyle t} t)^{-\beta/ \nu_{\scriptscriptstyle \parallel}}
{\tilde R}_{\scriptscriptstyle \text {pbc}} 
(0, 0, \infty, 1, \infty ) \, . \nonumber  
\end{eqnarray}
Comparing this result with Eq.\,(\ref{eq:rho_a_time_crit})
we get the scaling relation
\begin{equation}
\alpha \; = \; \frac{\, \beta \,}{\nu_{\scriptscriptstyle \parallel}} \, .
\label{eq:alpha_beta_nue}
\end{equation}
as well as ${\tilde R}_{\scriptscriptstyle \text {pbc}} 
(0, 0, \infty, 1, \infty )=1$.
But as can be seen from the inset
of Fig.\,\ref{fig:hyper_scal_01} this scaling relation
is clearly violated in $D=1$ and $D=2$.
For $D=3$ we think that the violation of Eq.\,(\ref{eq:alpha_beta_nue})
is hidden by the overlapping error-bars, i.e.,
the above scaling relation is
violated below the upper critical dimension,
as already observed in previous 
simulations~\cite{ROSSI_1}. 
Furthermore the violation of the scaling relation Eq.\,(\ref{eq:alpha_beta_nue})
explains why the well known hyperscaling relation
of directed percolation
\begin{equation}
\theta \, + \, \alpha \,+ \, \delta \; = \;
\frac{\,D\,}{z}  
\label{eq:theta_alpha_delta_D_z}
\end{equation}
is not fulfilled for absorbing phase transitions with a
conserved field.

It is worth to mention that this scaling anomaly is
not caused by the particular initial configuration
(random distribution of particles).
We have observed the same behavior for a 
{\it more} natural initial configuration,
where the correlations of the active and non-active
sites are not trivial.
In this case we start the simulations from a steady
state at the critical density with non-zero field~$h$.
Switching off the external field we have measured the
relaxation of the order parameter from the initial density 
$\rho_{\scriptscriptstyle \text {a},0}=\rho_{\scriptscriptstyle \text
a}(\rho_{\scriptscriptstyle \text c}, h)$.
For $\rho_{\scriptscriptstyle \text {a},0} \approx 0.1$
we observe the same scaling function and the same 
exponent $\alpha$ as in the case of a 
random initial configuration.

\begin{figure}[b]
  \includegraphics[width=8.0cm,angle=0]{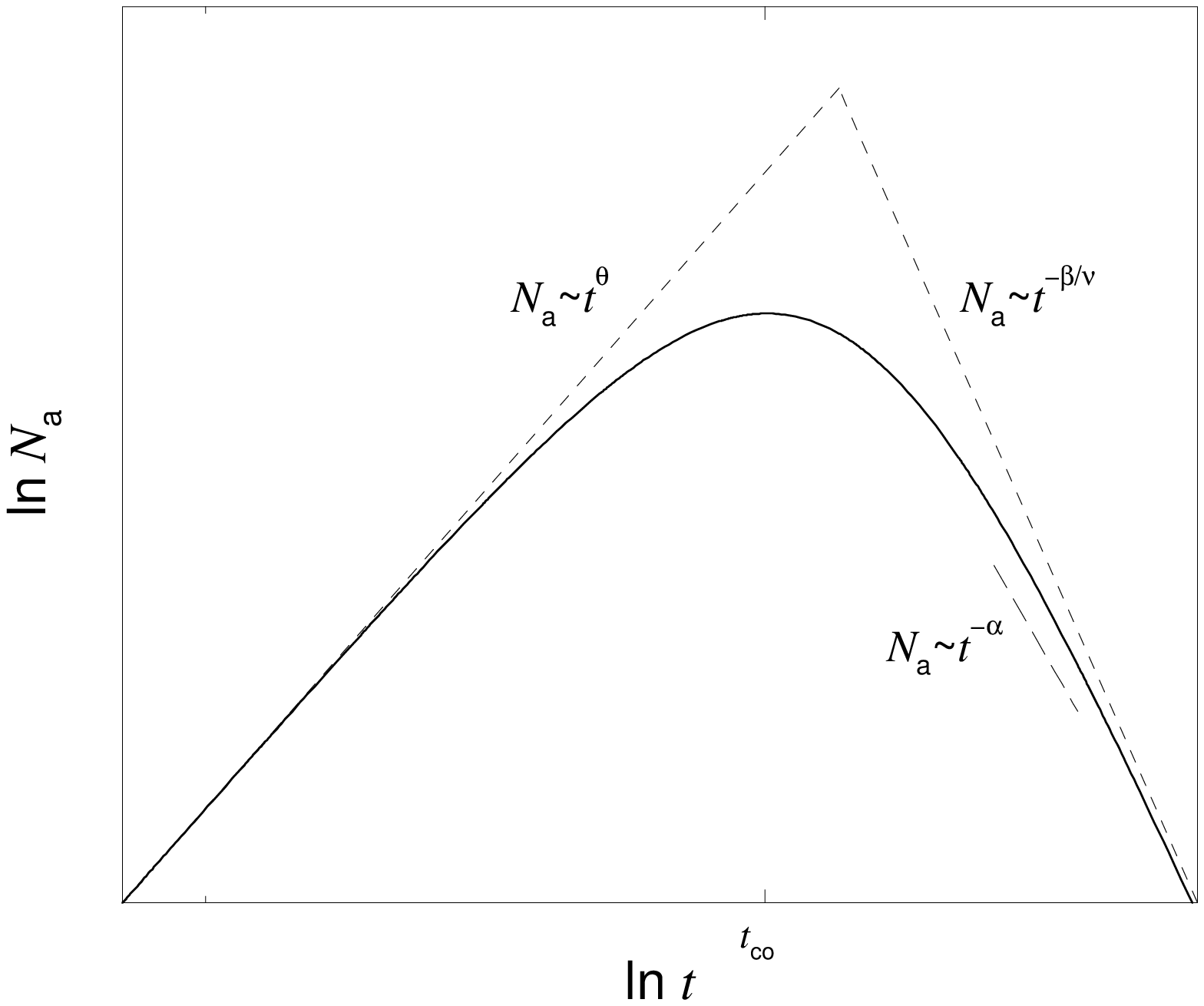}
  \caption{
  Sketch of the crossover behavior of the number of active
  sites~$N_{\scriptscriptstyle \text a}$.
  For $t \ll t_{\scriptscriptstyle \text {co}}$
  $N_{\scriptscriptstyle \text a}$ scales as $t^{\theta}$
  whereas it is expected to decrease for 
  $t \gg t_{\scriptscriptstyle \text {co}}$ as
  $N_{\scriptscriptstyle \text a} \propto t^{-\beta / 
  \nu_{\scriptscriptstyle \parallel}}$.
  It is possible that a too low exponent
  ($\alpha < \beta/\nu_{\scriptscriptstyle \parallel}$) 
  is observed in simulations if one has not
  reached the asymptotic scaling regime (long-dashed line). }
  \label{fig:sketch_crossover_tco}
\end{figure}

A possible explanation of the above scaling anomaly is that
the asymptotic scaling regime is so far not observed.
The situation is sketched in Fig.\,\ref{fig:sketch_crossover_tco}.
In the case of the two-dimensional CTTP 
the crossover takes place at 
$t_{\scriptscriptstyle \text {co}} \approx 1.14$
whereas finite-size effects occur at 
$t_{\scriptscriptstyle \text {FSS}} \approx 6000$
for $L=512$.
Thus the condition $t_{\scriptscriptstyle \text {co}} \ll t
\ll t_{\scriptscriptstyle \text {FSS}}$ seems to be 
fulfilled and one would expect to observe the
asymptotic scaling behavior $t^{-\beta/\nu_{\scriptscriptstyle ||}}$.
But it is known that crossovers span several
decades of magnitudes, usually 5, 6, 7  or even more decades 
(see for instance~\cite{LUEB_29} for a recent
work on crossover effects for the same universality 
class).
In this way it is possible that one observes in simulations a 
smaller exponent ($\alpha = 0.419$) than the asymptotic
value ($\beta/\nu_{\scriptscriptstyle ||}=0.522$). 
Future work on larger lattice sizes ($L\gg 512$) are
needed to clarify whether the scaling anomaly can be 
explained by a simple crossover effect.

Finally we consider the percolation probability 
$P_{\scriptscriptstyle \text {perc}}$ that a path
of active sites propagates through the system.
Obviously the percolation probability is related
to the survival probability via
\begin{equation}
P_{\scriptscriptstyle \text {perc}} \;  = \;
\lim_{t \to \infty} 
P_{\scriptscriptstyle \text {a}} (t) \, .
\label{eq:perc_prop_sur_prob}
\end{equation}
The percolation probability is expected to vanish
at the critical density according to
\begin{equation}
P_{\scriptscriptstyle \text {perc}} \;  \propto \;
\delta\rho^{\beta^{\prime}} \, .
\label{eq:perc_prop_beta}
\end{equation}
Assuming that the survival probability obeys the scaling form
\begin{eqnarray}
\label{eq:P_sur_scal}
& & a_{\scriptscriptstyle P} \, 
P_{\scriptscriptstyle \text a}(\delta\rho,L,t) \\
& & \sim 
\lambda^{-\delta \nu_{\scriptscriptstyle \parallel}} \,
{\tilde P}_{\scriptscriptstyle \text {pbc}}
(a_{\scriptscriptstyle \rho} \delta\rho \lambda ,
a_{\scriptscriptstyle t}  t \,
\lambda^{- \nu_{\scriptscriptstyle \parallel}} , 
a_{\scriptscriptstyle L} L
\lambda^{- \nu_{\scriptscriptstyle \perp}} )  \nonumber
\end{eqnarray}
we find in the thermodynamic limit 
\begin{eqnarray}
P_{\scriptscriptstyle \text {perc}} &  = &
\lim_{t \to \infty} 
P_{\scriptscriptstyle \text {a}} (\delta\rho, t, \infty) 
\nonumber \\
& \sim & a_{\scriptscriptstyle P}^{-1} \, 
(a_{\scriptscriptstyle \rho} \delta\rho)^{\delta
\nu_{\scriptscriptstyle \parallel}} \, {\tilde P}_{\scriptscriptstyle \text {pbc}}
(1,\infty , \infty )
\label{eq:p_perc_scal}
\end{eqnarray}
leading to the scaling relation 
\begin{equation}
\delta \; = \; \frac{\, \beta^{\prime}\,}{\nu_{\scriptscriptstyle \parallel}} \, .
\label{eq:delta_beta_nue}
\end{equation}
Thus the hyperscaling relations 
Eq.\,(\ref{eq:sigma_D_nu_perp_nu_par_nu_par_delta},
\ref{eq:theta_D_z_delta_beta_nu_par})
read now
\begin{eqnarray}
\label{eq:sigma_D_nu_perp_nu_par_betap}    
\sigma & = &
\nu_{\scriptscriptstyle \perp} D \, + \,
\nu_{\scriptscriptstyle \parallel} \, -  \, \beta^{\prime} \, , \\[3mm]
\label{eq:theta_D_z_betap_nu_par_beta_nu_par}
\theta & = &
\frac{\, \nu_{\scriptscriptstyle \perp} D\,
- \, \beta \, - \, \beta^{\prime} \,}{\nu_{\scriptscriptstyle \parallel}}  \, .
\end{eqnarray}

In the case of directed percolation 
$\beta$ equals $\beta^{\prime}$ due to 
a special symmetry under time reversal~\cite{GRASSBERGER_4}. 
In a more general context one expects that 
both exponents differ~\cite{HINRICHSEN_1}, 
for instance in systems with infinitely many absorbing states or
as in our case, in systems where
a conserved field couples to the order parameter.
The number of independent critical exponents
is therefore expected to be four (e.g.~$\beta$, $\beta^{\prime}$, 
$\nu_{\scriptscriptstyle \parallel}$, $\nu_{\scriptscriptstyle \perp}$) 
instead of three independent exponents for directed percolation 
($\beta$, $\nu_{\scriptscriptstyle \parallel}$, 
$\nu_{\scriptscriptstyle \perp}$).
In order to check this scenario we compare in the
inset of Fig.\,\ref{fig:hyper_scal_01} the spreading
exponent~$\delta$ with $\beta/\nu_{\scriptscriptstyle \parallel}$.
Surprisingly we observe that both values agree 
within the error-bars for all dimensions, suggesting 
$\beta = \beta^{\prime}$.
It is possible that the uncertainty of our estimates
hides a tiny difference of both exponents. 
A more accurate test of the scaling relation
$\beta = \beta^{\prime}$ could be obtained by a direct
measurement of the percolation probability which 
remains the topic of future research.

\section{Conclusions}

\label{sec:conc}
\begin{table}[t]
\caption{The non-universal metric factors
of the considered models.
The uncertainty of the metric factors is less than 5\%.}
\label{table:metric_factors}
\begin{tabular}{llll}
$$       &  $D=1$	& $D=2$	& $D=3$ \\  
\tableline 
CTTP$\quad\quad$ \\
$a_{\scriptscriptstyle \rho}$     &  $0.607\quad\quad$  & $0.341\quad\quad$	& $0.384$ \\
$a_{\scriptscriptstyle h}$        &  $0.220$	  & $0.013$	& $0.093$  \\
$a_{\scriptscriptstyle \Delta}$   &  $187.7$	  & $45.42$	& $24.51$  \\
$a_{\scriptscriptstyle L}$        &  $1014.$	  & $4.617$	& $2.173$  \\  
$a_{\scriptscriptstyle t}$        &  $1379.$	  & $24.90$	& $4.239$  \\  
$a_{\scriptscriptstyle P}$        &  $0.107$	  & $0.078$	& $0.094$  \\  
$a_{\scriptscriptstyle N}$        &  $6.062$	  & $2.818$	& $1.069$  \\
\tableline 
CLG$\quad\quad$ \\
$a_{\scriptscriptstyle \rho}$     &  $$       	  & $0.509$	& $0.434$  \\
$a_{\scriptscriptstyle h}$        &  $$	  	  & $0.062$	& $0.391$  \\
$a_{\scriptscriptstyle \Delta}$   &  $$	  	  & $9.241$	& $8.881$  \\
$a_{\scriptscriptstyle L}$        &  $$	  	  & $2.107$	& $1.441$  \\  
$a_{\scriptscriptstyle t}$        &  $$	  	  & $11.22$	& $3.140$  \\  
$a_{\scriptscriptstyle P}$        &  $$		  & $0.157$	& $0.183$  \\  
$a_{\scriptscriptstyle N}$        &  $$		  & $1.249$	& $0.569$  \\
\tableline 
Manna$\quad\quad$ \\
$a_{\scriptscriptstyle \rho}$     &  $0.662$       & $0.211$	& $0.311$  \\
$a_{\scriptscriptstyle h}$        &  $7.52\,10^{-5}$	  & $0.007$	& $0.074$  \\
$a_{\scriptscriptstyle \Delta}$   &  $588.9$	  & $78.56$	& $32.24$  \\
$a_{\scriptscriptstyle L}$        &  $205.9$	  & $6.011$	& $2.367$  \\  
$a_{\scriptscriptstyle t}$        &  $7.99\,10^{4}$	  & $35.53$	& $4.824$  \\  
$a_{\scriptscriptstyle P}$        &  $0.063$	  & $0.059$	& $0.089$  \\  
$a_{\scriptscriptstyle N}$        &  $64.02$	  & $3.600$	& $1.229$  
\end{tabular}
\end{table}

We analyzed numerically the critical behavior of 
three different models exhibiting a continuous
phase transition into an absorbing state.
In particular we introduce a method which allows
to consider finite-size effects in the steady state.
It is therefore possible to obtain accurate 
estimates of the correlation length exponent.
Additionally we determine the spreading exponents
which describe the spreading of activity
at the critical point.
A detailed analysis of certain scaling relations
shows that usual hyperscaling relations are fulfilled.
Only the activity spreading from a
homogeneous particle source exhibits a scaling
anomaly.
So far this scaling anomaly is not understood
and remains the topic of future research.
The number of independent exponents is due to the
scaling anomaly of the exponent $\alpha$ at least
four. 
In the case that $\beta \neq \beta^{\prime}$ it is
even five.

Since the hyperscaling relations 
Eqs.\,(\ref{eq:sigma_D_nu_perp_nu_par_nu_par_delta},
\ref{eq:theta_D_z_delta_beta_nu_par})
are fulfilled it is possible to connect the SOC avalanche
exponents to the steady state exponents of the corresponding
absorbing phase transition:
\begin{eqnarray}
\label{eq:soc_exp_apt_exp_tau_r}
\tau_{\scriptscriptstyle r} & = & 1 + \,z\, + D 
\, - \, \frac{\sigma}{\,\nu_{\scriptscriptstyle \perp}\,} \nonumber \\
& = & 1 \, + \, \frac{\beta^{\prime}}{\nu_{\scriptscriptstyle \perp}}, \\
\label{eq:soc_exp_apt_exp_tau_t}
\tau_{\scriptscriptstyle t} & = & 2 + \, \frac{\,D\,}{z} 
\, - \, \frac{\sigma}{\, \nu_{\scriptscriptstyle \parallel}\,} \nonumber \\
& = & 1 \, + \, \frac{\beta^{\prime}}{\nu_{\scriptscriptstyle \parallel}}, \\
\label{eq:soc_exp_apt_exp_tau_a}
\tau_{\scriptscriptstyle a} & = & 2 + \, \frac{\,z\,}{D} 
\, - \, \frac{\sigma}{\, D \nu_{\scriptscriptstyle \perp}\,} \nonumber \\
& = & 1 \, + \, \frac{\beta^{\prime}}{D \nu_{\scriptscriptstyle \perp}}, \\
\label{eq:soc_exp_apt_exp_tau_s}
\tau_{\scriptscriptstyle s} & = & 1 \, + \,  
\frac{ \,
\nu_{\scriptscriptstyle \parallel} 
+ \nu_{\scriptscriptstyle \perp} D - \sigma \,
}
{\nu_{\scriptscriptstyle \parallel} 
+ \nu_{\scriptscriptstyle \perp} D - \beta
} \nonumber \\ 
& = & 1 \, + \, \frac{   \beta^{\prime} }
{\, \nu_{\scriptscriptstyle \parallel} 
+ \nu_{\scriptscriptstyle \perp} D - \beta \,
}\, . 
\end{eqnarray}
In Fig.\,\ref{fig:soc_scal_01} we compare these values
with the avalanche exponents obtained from 
SOC simulations of the Manna model~\cite{LUEB_PHD,LUEB_9,DICKMAN_6}.
All SOC exponents agree within the error-bars with 
the avalanche exponents derived via the above scaling laws.
Thus it is possible to express the avalanche exponents
($\tau_{\scriptscriptstyle s}$, $\tau_{\scriptscriptstyle a}$,$\ldots$) 
of SOC systems in terms of the usual critical exponents
of a second order phase transition
($\beta$, $\nu_{\scriptscriptstyle \perp}$,
$\nu_{\scriptscriptstyle \parallel}$,
$\dots$).
In this way, the critical state of SOC sandpile models
is closely related to the critical state of an 
ordinary second order phase transition.

\acknowledgments
We would like to thank A.~Hucht, D.~Dhar, P.\,K.~Mohanty,
and A.~Vespignani
for helpful discussions and useful comments. 
P.\,C.~Heger thanks the Weizmann Institute for warm hospitality
during a visit when this work was in progress.
This work was financially supported by the 
Minerva Foundation (Max Planck Gesellschaft)
as well as by the 
Deutscher Akademischer Austauschdienst (DAAD).

\end{document}